\documentclass[nofootinbib,prl,superscriptaddress,a4paper,twocolumn]{revtex4-1}
\usepackage{geometry}
\usepackage[english]{babel}
\usepackage[latin1]{inputenc}
\usepackage{hyperref}
\usepackage{amsmath, amssymb, amsfonts, mathrsfs}
\usepackage{graphicx}
\usepackage{braket}

\usepackage{tikz}
\usetikzlibrary{calc}
\usepackage[all]{xy}

\newcommand{\Z}{\mathbb{Z}}

\tikzstyle WL=[line width=3pt,opacity=1.0]
\begin{document} 

\title{Quantized Graphs and Quantum Error Correction\\ {\small Dedicated to the $80^{th}$ Birthday of Arthur Jaffe}}

\author{Zhengwei Liu}
\email{liuzhengwei@mail.tsinghua.edu.cn}
\affiliation{Yau Mathematical Science Center and Department of Mathematics, Tsinghua University, Beijing 100084, China}

\begin{abstract}
Graph theory is important in information theory. We introduce a quantization process on graphs and apply the quantized graphs in quantum information. The quon language provides a mathematical theory to study such quantized graphs in a general framework. We give a new method to construct graphical quantum error correcting codes on quantized graphs and characterize all optimal ones.
We establish a further connection to geometric group theory and construct quantum low-density parity-check stabilizer codes on the Cayley graphs of groups. Their logical qubits can be encoded by the ground states of newly constructed exactly solvable models with translation-invariant local Hamiltonians. Moreover, the Hamiltonian is gapped in the large limit when the underlying group is infinite.
\end{abstract}

\date{\today}
\maketitle

\section{Introduction}

Graph theory has been widely used in classical information theory. Therefore it is natural to quantize these pictorial ideas and to apply them to quantum information. Our first main result is to introduce a quantization process on the graphs, that is compatible with the design of quantum error-correcting codes (QECC) and exactly solvable models. 

We give a new method to construct graphical QECC on any connected 4-valent graph, which we consider as a quantized graph. The graphical QECC is low-density parity-check (LDPC), when the quantized graph is a Cayley graph of a group. The logical qubits can be implemented as the ground states of a translation-invariant, local, gapped Hamiltonian. 
Several concepts for classical error corrections can be naturally quantized and captured by the quantized graphs in a pictorial way. In particular, we introduce a quantum linear system to characterized logical qubits as its solutions.
The quon language \cite{LWJ17quon} provides a mathematical picture language \cite{JL18} to study these quantized graphs. It illustrates internal symmetries and uncovers new dualities for QECC. We explain the quantization process and related concepts through our this paper using concrete examples. We state our results for the general cases. 
We plan to address the mathematical definitions and proofs in a companion paper~\cite{Liu-Companion}.

For example, let us start with the graph of the classical repetition code; then the quantized graph encodes Shor's quantum repetition code \cite{Shor95}, see the Section {\it SHOR'S QECC}. 
If we start with Kitaev's toric code \cite{Kit03}, possibly with line defect \cite{KitKon12} on a square lattice, then the quantized graph illustrates the encoding map, stabilizers, topological order, ground states and excitations in a pictorial way. The duality between vertices and plaquettes of the square lattice becomes a self-duality of the quantized graphs. In particular, the vertex operators and plaquette operators are unified as {\it cycle operators} on the quantized graphs, see the Section {\it TORIC CODES}. 

We can implement the stabilizers as cycle operators on quantized graphs not only for the toric code, but also for various kinds of stabilizer codes \cite{Got97}, 
For example, the optimal QECC for one logical qubit and one qubit error has 5 physical qubits, 1 logical qubits and distance 3, namely a [[5,1,3]] code \cite{LMPZ96,BDSW96}.
In the Section {\it GRAPHICAL QECC}, we recover this QECC using the quantized graph $K_5$, the complete graph with 5 vertices. We give the first pictorial interpretation of its stabilizer group as even-length cycles on the graph $K_5$. Furthermore, we discover a permutation group symmetry $S_5$ from the the quantized graph $K_5$, beyond the dihedral group symmetry $D_5$ of the graph state of this code.

In general, we consider a 4-valent connected graph as a quantized graph $\Gamma$, which may have no prequatization. The $n$ vertices of $\Gamma$ correspond to $n$ physical qubits.
The edges $E(\Gamma)$ of $\Gamma$ have no real physical interpretation in quantum information. We consider the edges as virtual concepts.
In the Section {\it {\it GRAPHICAL QECC}}, we construct graphical QECC on $\Gamma$ using these virtual concepts.
Take the Hilbert space $\ell^2(E(\gamma),\mathbb{F}_2)$ as maps from the edges to the field $\mathbb{F}_2=\{ 0,1\}$.
We obtain two additive subgroups, neutrally {\it charged graphs} $\mathcal{A}$ and cycles $\mathcal{B}$, which are dual to each other w.r.t. the inner product (modulo a {\it bipartite equivalence of $\mathcal{A}$}). 

In the quon language, every neutrally {\it charged graph} defines a $n$-qubit state. The states of charged graphs form an orthonormal basis of the n-qubit space, which we call the Fourier basis. 
Every cycle defines a cycle operator and all cycle operators form a stabilizer group.  For any subset $\mathcal{C}<\mathcal{A}$ or $\mathcal{L}<\mathcal{B}$, we construct a graphical QECC $(\Gamma,\mathcal{C})$ or $(\Gamma,\mathcal{L})$. They are the same QECC, if $\mathcal{C}=\mathcal{L}^{\perp}$ and $\mathcal{L}=\mathcal{C}^{\perp}$, see the Section {\it DUALITY FOR GRAPHICAL QECC}. 

The set $\mathcal{C}$ corresponds to logical qubits of the graphical QECC. 
We introduce the notion of a quantum linear system in the Section {\it QUANTUM LINEAR SYSTEMS}, and we use its solution to characterize logical qubits.

Moreover, we observe that computing the distance of a graphical QECC reduces to two classical problems on $\Gamma$, see the Section {\it DISTANCE}. This phenomenon is similar to the CSS code \cite{CS96,Ste96}.

Inspired by this observation, we characterize all optimal [[n,k,d]] graphic QECC on $\Gamma$ in the Section {\it FUNDAMENTAL THEOREMS}, where optimal means that neither $k$ nor $b$ can be larger without reducing the other.
As an application, we give several existence theorems. In summary, given any quantized graph, we can construct many graphical QECC on it.

Furthermore, we encode group symmetries in the design of graphical QECC. When the quantized graph is given by the Cayley graph of a group, we can encode additional group symmetries and locality in our graphical QECC. 
Given a group $\tilde{G}$ generated by $a$ and $b$ and a finite quotient $G$, we construct a graphical QECC $(\tilde{G},G,a,b)$ on the Cayley graph $\Gamma$ of $G$.
In particular, several QECC with geometric properties could be unified by the geometric group $\tilde{G}=<a,b : a^{p}, b^{q}, (ab)^{r}>$. 
Moreover, if $\tilde{G}$ is finitely presented and the length of the relations are much smaller than the order of $G$, which is usually the case, then the graphical QECC $(\tilde{G},G,a,b)$ is low-density parity-check (LDPC) and the coefficient matrix of its quantum linear system is sparse. We refer the readers to \cite{MMM04} for further discussions of LDPC codes.

For each $(\tilde{G},G,a,b)$, we construct an exactly solvable model on $\Gamma$ with a translation-invariant local Hamiltonian, such that its ground states are the logical qubits of $(\tilde{G},G,a,b)$. This correspondence should help to implement such QECC in the laboratory. An eigenbasis of the Hamiltonian is given by the states of charged graphs. If $\tilde{G}$ is an infinite group, and $G_n$ is a net of finite quotients of $G$, then we obtain a gapped Hamiltonian in the large scale limit.

\section{Graphs in Error-Correcting Codes}

\begin{figure}[h]
\begin{tikzpicture}
\foreach \x in {1,2,3}
{
\coordinate (A\x) at ({cos(\x*120)}, {sin(\x*120)});
\node at (A\x) {$\bullet$};
\node at ({.8*cos(\x*120+180)}, {.8*sin(\x*120+180)}) {\x};
}
\draw (A3)--(A1)--(A2)--(A3);
\end{tikzpicture}
\caption{graphical Representation of ECC}\label{Fig: RC 1}
\end{figure}
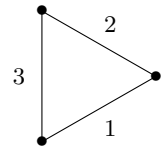

The repetition code is a classical Error-Correcting Code (ECC). It encodes one logical bit by three physical bits, and corrects the error of one bit flip as shown in Table~\ref{Table: Repetition Code}. 
\begin{table}[h]
\begin{tabular}{|cc|c|c|}
\hline 
Input &(logical bits) & 0 & 1\\
\hline 
Encoding map &(physical bits) & 000 & 111 \\
\hline 
Error & (bit flip)& 00\textcolor{red}{1} & 11\textcolor{red}{0} \\
\hline 
Error Correction & (majority) & 0 & 1 \\
\hline 
\end{tabular}
\caption{Repetition Code}\label{Table: Repetition Code}
\end{table}
The image of the encoding map can be described through the solution of a linear system $A\eta=0$ over $\mathbb{F}_2$, the field with two elements $\{0,1\}$, where
\begin{align}\label{Equ: linear system}
A=\left[
\begin{array}{ccc}
1 & 1 & 0 \\
0 & 1 & 1 \\
1 & 0 & 1 \\
\end{array}
\right]
\end{align}
One can represent the matrix $A$ by a graph $\Psi$ as in Fig.~\ref{Fig: RC 1}.
Each edge corresponds to a bit and each vertex corresponds to a relation that the sum of bits on the adjacent edges is zero mod 2.

Such a graphical representation is important in the study of classical error corrections in information theory through graph theory. In particular, graphs with large girth, namely the minimal length of cycles, usually give ECC with large distance. Expander graphs are such graphs.

\section{Quantization of Graphs}
\begin{figure}[h]
\raisebox{-1cm}{
\begin{tikzpicture}
\foreach \x in {1,2,3}
{
\coordinate (A\x) at ({cos(\x*120)}, {sin(\x*120)});
\node at (A\x) {$\bullet$};
\node at ({.8*cos(\x*120+180)}, {.8*sin(\x*120+180)}) {\x};
}
\draw (A3)--(A1)--(A2)--(A3);
\end{tikzpicture}}
$\longrightarrow$
\raisebox{-1cm}{
\begin{tikzpicture}
\foreach \x in {1,2,3}
{
\coordinate (B\x) at ({.57*cos(\x*120+180)}, {.57*sin(\x*120+180)});
\node at ({.8*cos(\x*120+180)}, {.8*sin(\x*120+180)}) {\x};
}
\fill[gray!50] (B1) to [bend right=30] (B2) to [bend left=90] (B1);
\draw (B1) to [bend right=30] (B2) to [bend left=90] (B1);
\fill[gray!50] (B2) to [bend right=30] (B3) to [bend left=90] (B2);
\draw (B2) to [bend right=30] (B3) to [bend left=90] (B2);
\fill[gray!50] (B3) to [bend right=30] (B1) to [bend left=90] (B3);
\draw (B3) to [bend right=30] (B1) to [bend left=90] (B3);
\foreach \x in {1,2,3}
{
\fill[white] (B\x) circle (.1);
\draw (B\x) circle (.1);
}
\end{tikzpicture}}
\caption{Quantized Graph for the Repetition Code}\label{Fig: Quantized RC}
\end{figure}
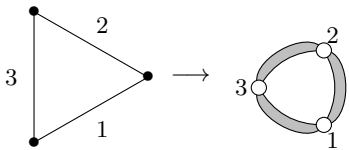

We give a quantization process on graphs as shown in Table~\ref{Table: Quantization}. For example, the quantized graph $\overline{\Psi}$ of $\Psi$ is illustrated in Fig.~\ref{Fig: Quantized RC}. We can de-quantize the graph $\overline{\Psi}$ by reversing the process. Note that if we de-quantize the graph $\overline{\Psi}$ after switching the alternating shadings of its regions, and then we end up with the dual graph of $\Psi$. This quantization process naturally captures the duality of the graphs. On the quantized graph, the vertex corresponds to the bit, the shaded region corresponds to the relation. It is less obvious that the unshaded region corresponds to the logic qubit. We use the quon language \cite{LWJ17quon} to explain this phenomenon and its application in quantum information.
\begin{table}[h]
\begin{tabular}{c|c|c|c}
 graph $\Psi$ & quantization & quantized graph $\overline{\Psi}$ & ECC\\
\hline
edge & $\longrightarrow$ & 4-valent vertex (disc) & bit \\
vertex & $\longrightarrow$ & shaded region & relation \\
region & $\longrightarrow$ & unshaded region & logical bit
\end{tabular}
\caption{Quantization of Graphs}\label{Table: Quantization}
\end{table}

\section{Quon Language}

The quon language is a 3D picture language originally designed for quantum information \cite{LWJ17quon}. The pictures are given by braided charged strings in three manifolds in the three dimensional space. The pictures in the bulk represent gates projectively and the pictures on the boundary represent states linearly.  We recall some relevant properties of the quon language, which we are applying for quantum error corrections in this paper. The braid satisfies Reidemeister moves of type I, II, III. The charge behaves like a Majorana fermion, which we represent by a dot ``$\bullet$'' in the quon language for qubits: 
\begin{center}
\begin{tabular}{ll}
\begin{tikzpicture}
\begin{scope}[scale=.5]
\draw (0,0) circle (.5);
\node at (1.8,0) {$=\sqrt{2}$};
\end{scope}
\end{tikzpicture},
&
\begin{tikzpicture}
\begin{scope}[scale=.5]
\draw (0,0) circle (.5);
\node at (1.6,0) {$=0$};
\fill (-.5,0) circle (.1);
\end{scope}
\end{tikzpicture}, 
\\
\; 
\begin{tikzpicture}
\begin{scope}[scale=.5]
\draw (0,0)--(0,1);
\fill (0,.3) circle (.1);
\fill (0,.7) circle (.1);
\node at (1,.5) {$=$};
\draw (2,0)--++(0,1);
\end{scope}
\end{tikzpicture} \; , 
&
\begin{tikzpicture}
\begin{scope}[scale=.5]
\draw (0,-.2)--(0,0) arc (180:0:.5)--++(0,-.2);
\fill (0,0) circle (.1);
\node at (1.75,.15) {$=i$};
\draw (2.5,-.2)--(2.5,0) arc (180:0:.5)--++(0,-.2);
\fill (3.5,0) circle (.1);
\end{scope}
\end{tikzpicture},
\\
\begin{tikzpicture}
\begin{scope}[scale=.5]
\draw (1,0) -- (0,1);
\draw[white,WL] (0,0) -- (1,1);
\draw (0,0) -- (1,1);
\fill (.8,.2) circle (.1);
\node at (1.75,.5) {$=i$};
\begin{scope}[shift={(2.5,0)}]
\draw (1,0) -- (0,1);
\draw[white,WL] (0,0) -- (1,1);
\draw (0,0) -- (1,1);
\fill (.2,.8) circle (.1);
\end{scope}
\end{scope}
\end{tikzpicture},
&
\begin{tikzpicture}
\begin{scope}[scale=.5]
\draw (0,0)--(.75,.25) --(.25,.75) --(1,1);
\draw[white,WL] (.25,.25)--(.75,.75) ;
\draw (0,1) --(.25,.25)--(.75,.75) --(1,0);
\node at (1.75,.5) {$=$};
\begin{scope}[shift={(2.5,0)}]
\draw (0,0) -- (1,1);
\draw[white,WL] (1,0) -- (0,1);
\draw (1,0) -- (0,1);
\end{scope}
\end{scope}
\end{tikzpicture},
\\
\begin{tikzpicture}
\begin{scope}[scale=.5]
\draw (0,0)--(0,1);
\draw (1,0)--(1,1);
\fill (0,.3) circle (.1);
\fill (1,.7) circle (.1);
\node at (2,.5) {$=-$};
\begin{scope}[shift={(3,0)}]
\draw (0,0)--(0,1);
\draw (1,0)--(1,1);
\fill (1,.3) circle (.1);
\fill (0,.7) circle (.1);
\end{scope}
\node at (5,.5) {$=-i$};
\begin{scope}[shift={(6.5,0)}]
\draw (0,0)--(0,1);
\draw (1,0)--(1,1);
\fill (1,.5) circle (.1);
\fill (0,.5) circle (.1);
\end{scope}
\end{scope}
\end{tikzpicture} \; ;
&
\\
{\color{black}
\begin{tikzpicture}
\begin{scope}[scale=.5]
\draw (0,0) -- (1,1);
\draw[white,WL] (1,0) -- (0,1);
\draw (1,0) -- (0,1);
\node at (1.75,.5) {$=$};
\begin{scope}[shift={(2.5,0)}]
\draw (1,0) -- (0,1);
\draw[white,WL] (0,0) -- (1,1);
\draw (0,0) -- (1,1);
\fill (.8,.2) circle (.1);
\fill (.2,.2) circle (.1);
\end{scope}
\end{scope}
\end{tikzpicture},}
&
\begin{tikzpicture}
\begin{scope}[scale=.5]
\draw (1,0) -- (0,1);
\draw[white,WL] (0,0) -- (1,1);
\draw (0,0) -- (1,1);
\fill (.2,.2) circle (.1);
\node at (1.75,.5) {$=-$};
\begin{scope}[shift={(2.5,0)}]
\draw (1,0) -- (0,1);
\draw[white,WL] (0,0) -- (1,1);
\draw (0,0) -- (1,1);
\fill (.8,.8) circle (.1);
\end{scope}
\end{scope}
\end{tikzpicture}.
\end{tabular}
\end{center}
One should not confuse our notation for a charge with a vertex in a classical graph, such as in Fig.\ref{Fig: RC 1}.
The upper seven relations can be generalized to qudits as shown in  \cite{LWJ17quon}. 
The last two new relations hold only for qubits and they are crucial in our study of QECC. 
The strings and charges move freely in the 3D space after ignoring the global phase. Modulo the charges, we can switch the layers of the braid. This is important to do pictorial computation efficiently, while it is difficult to evaluate a link in polynomial time in general. Furthermore, this property implies that our construction of the {\it graphical QECC} is essentially independent of the choice of the layers of the braids.

The following string-genus relation removes a hole surrounded by a string. While it still has no physical interpretation, it turns out to be important in our study of QECC. Its pictorial representation is
\begin{center}
\begin{tikzpicture}
\begin{scope}[scale=.5]
\draw (0,0) circle (1);
\draw[blue] (-.5,.2) arc (-180:0:.5);
\draw[blue] (-.3,0) arc (180:0:.3);
\node at (2.5,0) {$\displaystyle=\frac{1}{\sqrt{2}}.$};
\end{scope}
\end{tikzpicture}
\end{center}
%
We represent the 1-qubit XYZ basis using the following diagrams in a hemisphere:
\begin{align*}
\sqrt{2}~\ket{0}_Z&=
\raisebox{-.0cm}{
\begin{tikzpicture}
\begin{scope}[scale=.5]
\draw (0,0) arc (180:0:.5);
\draw (2,0) arc (180:0:.5);
\end{scope}
\end{tikzpicture}} \\
\sqrt{2}~\ket{0}_Y&=
\raisebox{-.25cm}{
\begin{tikzpicture}
\begin{scope}[scale=.5]
\draw (0,0) arc (180:0:1);
\draw[white,WL] (1,0) arc (180:0:1);
\draw (1,0) arc (180:0:1);
\end{scope}
\end{tikzpicture}} \\
\sqrt{2}~\ket{0}_X&=
\raisebox{-.25cm}{
\begin{tikzpicture}
\begin{scope}[scale=.5]
\draw (0,0) arc (180:0:1.5);
\draw (1,0) arc (180:0:.5);
\end{scope}
\end{tikzpicture}}
\end{align*}
We obtain $\ket{1}_{Z}, \ket{1}_{Y}, \ket{1}_{X}$ by adding a pair of charges to the pair of strings of $\ket{0}_{Z}, \ket{0}_{Y}, \ket{0}_{X}$, respectively.
We represent the 1-qubit Pauli X, Y, Z gates using the following diagrams in a cylinder:
\begin{align*}
I&=
\raisebox{-.25cm}{
\begin{tikzpicture}
\begin{scope}[scale=.5]
\foreach \x in {0,1,2,3}
\draw (\x,0) --++ (0,1);
\end{scope}
\end{tikzpicture}}
\quad=
\raisebox{-.25cm}{
\begin{tikzpicture}
\begin{scope}[scale=.5]
\foreach \x in {0,1,2,3}
\draw (\x,0) --++ (0,1);
\fill (0,.5) circle (.1); 
\fill (1,.5) circle (.1); 
\fill (2,.5) circle (.1); 
\fill (3,.5) circle (.1); 
\end{scope}
\end{tikzpicture}}
 \\
Z&=
\raisebox{-.25cm}{
\begin{tikzpicture}
\begin{scope}[scale=.5]
\foreach \x in {0,1,2,3}
\draw (\x,0) --++ (0,1);
\fill (0,.5) circle (.1); 
\fill (1,.5) circle (.1); 
\end{scope}
\end{tikzpicture}}
\quad=
\raisebox{-.25cm}{
\begin{tikzpicture}
\begin{scope}[scale=.5]
\foreach \x in {0,1,2,3}
\draw (\x,0) --++ (0,1);
\fill (2,.5) circle (.1); 
\fill (3,.5) circle (.1); 
\end{scope}
\end{tikzpicture}}
 \\
Y&=
\raisebox{-.25cm}{
\begin{tikzpicture}
\begin{scope}[scale=.5]
\foreach \x in {0,1,2,3}
\draw (\x,0) --++ (0,1);
\fill (0,.5) circle (.1); 
\fill (2,.5) circle (.1); 
\end{scope}
\end{tikzpicture}}
\quad=
\raisebox{-.25cm}{
\begin{tikzpicture}
\begin{scope}[scale=.5]
\foreach \x in {0,1,2,3}
\draw (\x,0) --++ (0,1);
\fill (1,.5) circle (.1); 
\fill (3,.5) circle (.1); 
\end{scope}
\end{tikzpicture}} \\
X&=
\raisebox{-.25cm}{
\begin{tikzpicture}
\begin{scope}[scale=.5]
\foreach \x in {0,1,2,3}
\draw (\x,0) --++ (0,1);
\fill (0,.5) circle (.1); 
\fill (3,.5) circle (.1); 
\end{scope}
\end{tikzpicture}}
\quad=
\raisebox{-.25cm}{
\begin{tikzpicture}
\begin{scope}[scale=.5]
\foreach \x in {0,1,2,3}
\draw (\x,0) --++ (0,1);
\fill (1,.5) circle (.1); 
\fill (2,.5) circle (.1); 
\end{scope}
\end{tikzpicture}}
\end{align*}
Therefore, adding Pauli matrices to the states becomes adding charges to the graph.

\section{Shor's QECC}
In this section, we recall Shor's QECC in \cite{Shor95}. We first illustrate our pictorial concepts for these examples.
The quantum analogue of the repetition code was given by Shor.
The encoding map is 
\begin{align}\label{Equ: Encoding X}
\iota (\alpha \ket{0}+\beta \ket{1})=\alpha\ket{000}+\beta\ket{111} \; .
\end{align}
It encodes one logical qubit by three physical qubits, and corrects one Pauli $X$ error.

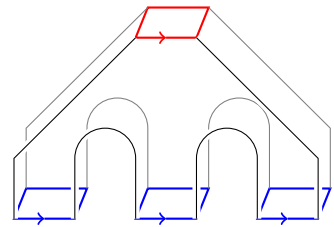
\begin{figure}[h]
\begin{tikzpicture}
\begin{scope}[scale=.8]
\draw[red,thick,->] (.3,6-.5)--++(.5,0)--++(.2,.5)--++(-1,0)--++(-.2,-.5)--++(.5,0);
\draw[blue,thick,->] (-2+.3,3-.5)--++(.5,0)--++(.2,.5)--++(-1,0)--++(-.2,-.5)--++(.5,0);
\draw[blue,thick,->] (+.3,3-.5)--++(.5,0)--++(.2,.5)--++(-1,0)--++(-.2,-.5)--++(.5,0);
\draw[blue,thick,->] (2+.3,3-.5)--++(.5,0)--++(.2,.5)--++(-1,0)--++(-.2,-.5)--++(.5,0);
\draw[gray] (-2,3)--++(0,1)--++(2,2);
\draw[gray] (3,3)--++(0,1)--++(-2,2);
\draw[gray] (0,3)--++(0,1) arc (0:180:.5)--++(0,-1);
\draw[gray] (2,3)--++(0,1) arc (0:180:.5)--++(0,-1);
\begin{scope}[shift={(-.2,-.5)}]
\draw[white,WL] (-2,3)--++(0,1)--++(2,2);
\draw[white,WL] (3,3)--++(0,1)--++(-2,2);
\draw[white,WL] (0,3)--++(0,1) arc (0:180:.5)--++(0,-1);
\draw[white,WL] (2,3)--++(0,1) arc (0:180:.5)--++(0,-1);
\end{scope}
\begin{scope}[shift={(-.2,-.5)}]
\draw (-2,3)--++(0,1)--++(2,2);
\draw (3,3)--++(0,1)--++(-2,2);
\draw (0,3)--++(0,1) arc (0:180:.5)--++(0,-1);
\draw (2,3)--++(0,1) arc (0:180:.5)--++(0,-1);
\end{scope}
\end{scope}
\end{tikzpicture}
\caption{Shor's Encoding Map in the Quon Language}\label{Fig: QRC}
\end{figure}

\begin{figure}[h]
$$
\raisebox{-3cm}{
\begin{tikzpicture}
\begin{scope}[scale=.8]
\draw[red,thick,->] (.3,6-.5)--++(.5,0)--++(.2,.5)--++(-1,0)--++(-.2,-.5)--++(.5,0);
\draw[blue,thick,->] (.3,0-.5)--++(.5,0)--++(.2,.5)--++(-1,0)--++(-.2,-.5)--++(.5,0);
\draw[gray] (0,0)--++(-2,2)--++(0,2)--++(2,2);
\draw[gray] (1,0)--++(2,2)--++(0,2)--++(-2,2);
\draw[gray] (-1,2) arc (-180:0:.5)--++(0,2) arc (0:180:.5)--++(0,-2);
\draw[gray] (1,2) arc (-180:0:.5)--++(0,2) arc (0:180:.5)--++(0,-2);
\begin{scope}[shift={(1.5,3)}, scale=.5]
\draw[blue] (-.5,.2) arc (-180:0:.5);
\draw[blue] (-.3,0) arc (180:0:.3);
\end{scope}
\begin{scope}[shift={(-.5,3)}, scale=.5]
\draw[blue] (-.5,.2) arc (-180:0:.5);
\draw[blue] (-.3,0) arc (180:0:.3);
\end{scope}
\begin{scope}[shift={(-.2,-.5)}]
\draw[white,WL] (0,0)--++(-2,2)--++(0,2)--++(2,2);
\draw[white,WL] (1,0)--++(2,2)--++(0,2)--++(-2,2);
\draw[white,WL] (-1,2) arc (-180:0:.5)--++(0,2) arc (0:180:.5)--++(0,-2);
\draw[white,WL] (1,2) arc (-180:0:.5)--++(0,2) arc (0:180:.5)--++(0,-2);
\end{scope}
\begin{scope}[shift={(-.2,-.5)}]
\draw (0,0)--++(-2,2)--++(0,2)--++(2,2);
\draw (1,0)--++(2,2)--++(0,2)--++(-2,2);
\draw (-1,2) arc (-180:0:.5)--++(0,2) arc (0:180:.5)--++(0,-2);
\draw (1,2) arc (-180:0:.5)--++(0,2) arc (0:180:.5)--++(0,-2);
\end{scope}
\begin{scope}[shift={(-2.3,2.2)}]
\fill[white] (0,0) rectangle (1.2,.6);
\draw (0,0) rectangle (1.2,.6);
\node at (.6,.3) {$E_p^{\dagger}$};
\end{scope}
\begin{scope}[shift={(2-2.3,2.2)}]
\fill[white] (0,0) rectangle (1.2,.6);
\draw (0,0) rectangle (1.2,.6);
\node at (.6,.3) {$E_q$};
\end{scope}
\end{scope}
\end{tikzpicture}}
=
\raisebox{-3cm}{
\begin{tikzpicture}
\begin{scope}[scale=.8]
\draw[red,thick,->] (.3,6-.5)--++(.5,0)--++(.2,.5)--++(-1,0)--++(-.2,-.5)--++(.5,0);
\draw[blue,thick,->] (.3,0-.5)--++(.5,0)--++(.2,.5)--++(-1,0)--++(-.2,-.5)--++(.5,0);
\draw[gray] (0,0)--++(0,6);
\draw[gray] (1,0)--++(0,6);
\begin{scope}[shift={(-.2,-.5)}]
\draw (0,0)--++(0,6);
\draw (1,0)--++(0,6);
\end{scope}
\begin{scope}[shift={(-.7,2.2)}]
\fill[white] (0,0) rectangle (.6,.6);
\draw (0,0) rectangle (.6,.6);
\node at (.3,.3) {$?$};
\end{scope}
\end{scope}
\end{tikzpicture}}
=C_{pq}
\raisebox{-3cm}{
\begin{tikzpicture}
\begin{scope}[scale=.8]
\draw[red,thick,->] (.3,6-.5)--++(.5,0)--++(.2,.5)--++(-1,0)--++(-.2,-.5)--++(.5,0);
\draw[blue,thick,->] (.3,0-.5)--++(.5,0)--++(.2,.5)--++(-1,0)--++(-.2,-.5)--++(.5,0);
\draw[gray] (0,0)--++(0,6);
\draw[gray] (1,0)--++(0,6);
\begin{scope}[shift={(-.2,-.5)}]
\draw (0,0)--++(0,6);
\draw (1,0)--++(0,6);
\end{scope}
\end{scope}
\end{tikzpicture}}
$$
\caption{Diagrammatic Verification of the Knill-Laffamme Condition: The 1-qubit error is either $I$ or $X$, so it is a blackbox defined on 2-strings, instead of 4-strings.  The first equality comes from the string-genus relation, which removes the two holes and surrounding strings. The second equality follows from the fact that the neutral diagram with two boundary points is a scalar multiple of a string.}\label{Fig: QRC LK}
\end{figure}
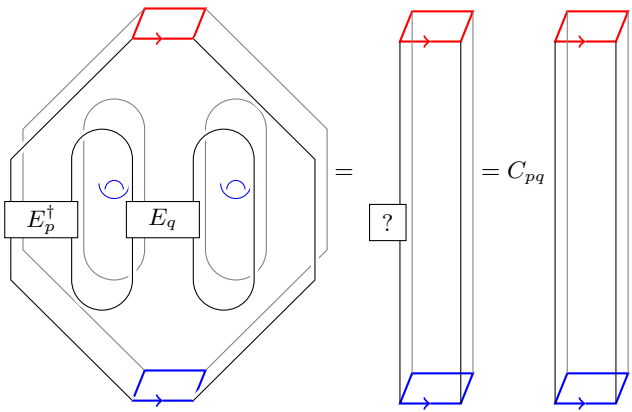

In the quon language, the encoding map $\iota$ can be represented by the tangle with one input disc and three output discs as shown in Fig.~\ref{Fig: QRC}. 
One can verify the detectable errors using the Knill-Laflamme condition \cite{KniLaf97}: For any $E_p, E_q \in \{I, X_1, X_2, X_3\}$, $\iota^{\dagger} E_p^\dagger E_{q} \iota =C_{pq}$, for some scalar $C_{pq}$. The diagrammatic verification without algebraic computations is shown in Fig.~\ref{Fig: QRC LK}. Essentially we want to show that the error will not affect the computation up to a scalar. The choice of two strings out of four to represent a Pauli $X$ in Fig.~\ref{Fig: QRC LK} is useful to design a good diagrammatic verification. The string-genus relation in the quon language \cite{LWJ17quon} is the key to change the shape of the surface in the diagrammatic verification. It replaces a genus with a surrounding string by a scalar $\frac{1}{\sqrt{2}}$.

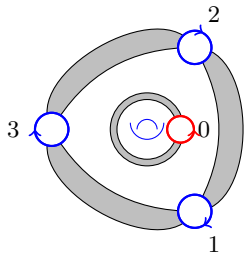
\begin{figure}[h]
\raisebox{-1.5cm}{
\begin{tikzpicture}
\begin{scope}[scale=2.2]
\foreach \x in {1,2,3}
{
\coordinate (B\x) at ({.57*cos(\x*120+180)}, {.57*sin(\x*120+180)});
\node at ({.8*cos(\x*120+180)}, {.8*sin(\x*120+180)}) {\x};
}
\fill[gray!50] (B1) to [bend right=30] (B2) to [bend left=90] (B1);
\draw (B1) to [bend right=30] (B2) to [bend left=90] (B1);
\fill[gray!50] (B2) to [bend right=30] (B3) to [bend left=90] (B2);
\draw (B2) to [bend right=30] (B3) to [bend left=90] (B2);
\fill[gray!50] (B3) to [bend right=30] (B1) to [bend left=90] (B3);
\draw (B3) to [bend right=30] (B1) to [bend left=90] (B3);
\foreach \x in {1,2,3}
{
\fill[white] (B\x) circle (.1);
\draw[blue,thick] (B\x) circle (.1);
\draw[blue,thick,<-]  ({.67*cos(\x*120+180)}, {.67*sin(\x*120+180)}) arc (\x*120-180:\x*120+180:.1);
}
\begin{scope}[scale=.2]
\node at (1.7,0) {0}; 
\fill[gray!50] (0,0) circle (1.1);
\fill[white] (0,0) circle (.9);
\draw (0,0) circle (1.1);
\draw (0,0) circle (.9);
\fill[white] (1,0) circle (.4);
\draw[red,thick] (1,0) circle (.4);
\draw[red,thick,->] (1.4,0) arc (0:360:.4);
\draw[blue] (-.5,.2) arc (-180:0:.5);
\draw[blue] (-.3,0) arc (180:0:.3);
\end{scope}
\end{scope}
\end{tikzpicture}}
\caption{Shor's Encoding Map by the Quantized Graph: It has one input disc labelled by 0, and three output discs labelled by 1, 2, 3.}\label{Fig: QRC Quon}
\end{figure}

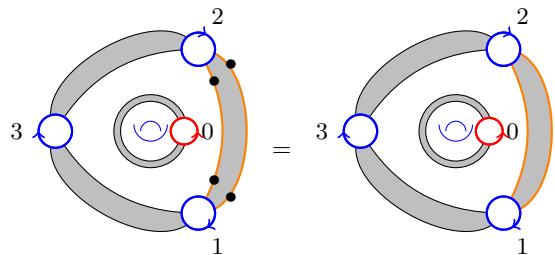
\begin{figure}[h]
\raisebox{-1.5cm}{
\begin{tikzpicture}
\begin{scope}[scale=2.2]
\foreach \x in {1,2,3}
{
\coordinate (B\x) at ({.57*cos(\x*120+180)}, {.57*sin(\x*120+180)});
\node at ({.8*cos(\x*120+180)}, {.8*sin(\x*120+180)}) {\x};
}
\fill[gray!50] (B1) to [bend right=30] (B2) to [bend left=90] (B1);
\draw[orange,thick] (B1) to [bend right=30] (B2) to [bend left=90] (B1);
\fill[gray!50] (B2) to [bend right=30] (B3) to [bend left=90] (B2);
\draw (B2) to [bend right=30] (B3) to [bend left=90] (B2);
\fill[gray!50] (B3) to [bend right=30] (B1) to [bend left=90] (B3);
\draw (B3) to [bend right=30] (B1) to [bend left=90] (B3);
\node at (.38,.3) {$\bullet$};
\node at (.48,.4) {$\bullet$};
\node at (.38,-.3) {$\bullet$};
\node at (.48,-.4) {$\bullet$};
\foreach \x in {1,2,3}
{
\fill[white] (B\x) circle (.1);
\draw[blue,thick] (B\x) circle (.1);
\draw[blue,thick,<-]  ({.67*cos(\x*120+180)}, {.67*sin(\x*120+180)}) arc (\x*120-180:\x*120+180:.1);
}
\begin{scope}[scale=.2]
\node at (1.7,0) {0}; 
\fill[gray!50] (0,0) circle (1.1);
\fill[white] (0,0) circle (.9);
\draw (0,0) circle (1.1);
\draw (0,0) circle (.9);
\fill[white] (1,0) circle (.4);
\draw[red,thick] (1,0) circle (.4);
\draw[red,thick,->] (1.4,0) arc (0:360:.4);
\draw[blue] (-.5,.2) arc (-180:0:.5);
\draw[blue] (-.3,0) arc (180:0:.3);
\end{scope}
\end{scope}
\end{tikzpicture}}
=
\raisebox{-1.5cm}{
\begin{tikzpicture}
\begin{scope}[scale=2.2]
\foreach \x in {1,2,3}
{
\coordinate (B\x) at ({.57*cos(\x*120+180)}, {.57*sin(\x*120+180)});
\node at ({.8*cos(\x*120+180)}, {.8*sin(\x*120+180)}) {\x};
}
\fill[gray!50] (B1) to [bend right=30] (B2) to [bend left=90] (B1);
\draw[orange,thick] (B1) to [bend right=30] (B2) to [bend left=90] (B1);
\fill[gray!50] (B2) to [bend right=30] (B3) to [bend left=90] (B2);
\draw (B2) to [bend right=30] (B3) to [bend left=90] (B2);
\fill[gray!50] (B3) to [bend right=30] (B1) to [bend left=90] (B3);
\draw (B3) to [bend right=30] (B1) to [bend left=90] (B3);
\foreach \x in {1,2,3}
{
\fill[white] (B\x) circle (.1);
\draw[blue,thick] (B\x) circle (.1);
\draw[blue,thick,<-]  ({.67*cos(\x*120+180)}, {.67*sin(\x*120+180)}) arc (\x*120-180:\x*120+180:.1);
}
\begin{scope}[scale=.2]
\node at (1.7,0) {0}; 
\fill[gray!50] (0,0) circle (1.1);
\fill[white] (0,0) circle (.9);
\draw (0,0) circle (1.1);
\draw (0,0) circle (.9);
\fill[white] (1,0) circle (.4);
\draw[red,thick] (1,0) circle (.4);
\draw[red,thick,->] (1.4,0) arc (0:360:.4);
\draw[blue] (-.5,.2) arc (-180:0:.5);
\draw[blue] (-.3,0) arc (180:0:.3);
\end{scope}
\end{scope}
\end{tikzpicture}}
\caption{Cycle Operator as Stabilizers: $\textcolor{orange}{O_L=Z_1Z_2}$}\label{Fig: Cycle Operator}
\end{figure}

Now we give a second diagrammatic representation of Shor's encoding map using the quantized graph $\overline{\Psi}$. 
We add a hole surrounded by an input disc in the unshaded region of $\overline{\Psi}$ as shown in Fig.~\ref{Fig: QRC Quon}. 
From this point of view, the unshaded region of $\overline{\Psi}$ corresponds to the logic bit (or qubit) as mentioned in Table~\ref{Table: Quantization}.
The image of Shor's encoding map is the invariant space of the stabilizers $Z_1Z_2$, $Z_2Z_3$, $Z_1Z_3$. One can read these Pauli-$Z$ stabilizers from the matrix $A$ in Equation \eqref{Equ: linear system} as well.
The quantized graph naturally captures the these stabilizers as the cycles on the boundary of the shaded regions.  
Precisely, let $L$ be the cycle on the boundary of the shaded region between the discs 1 and 2.
We define the {\it cycle operator} $O_L$ acting on the encoding map $\iota$, by adding pairs of charges on the cycle $L$ near by the discs 1 and 2.
By the diagrammatic Kitaev's map, $O_L=Z_1Z_2$. The cycle operator $O_L$ stabilizes the encoding map $\iota$, namely $O_L\iota=\iota$, because each edge in the cycle $L$ contains two changes, which will cancel each other as illustrated in Fig.~\ref{Fig: Cycle Operator}. 
Therefore the cycle operators are stabilizers.
We are going to extend the quantization process in Table~\ref{Table: Quantization} to quantum error-correcting codes, see Table~\ref{Table: Quantization QECC}.

\begin{table}[h]
\begin{tabular}{c|c|c|c}
 graph & quantization & quantized graph & QECC\\
\hline
edge & $\longrightarrow$ & 4-valent vertex (disc) & qubit \\
vertex & $\longrightarrow$ & (shaded) cycles & stabilizers \\
region & $\longrightarrow$ & unshaded region & logical qubit
\end{tabular}
\caption{Quantization of Graphs for QECC}\label{Table: Quantization QECC}
\end{table} 

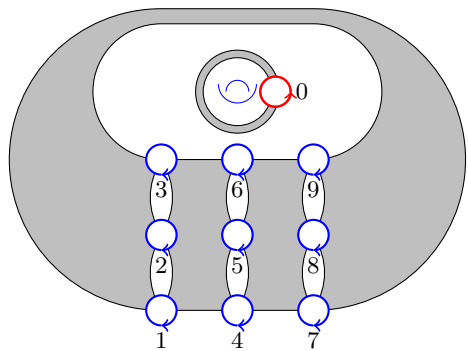
\begin{figure}[h]
\begin{tikzpicture}
\fill[gray!50] (0,0)--++(2,0) arc (-90:90:2) --++(-2,0) arc (90:270:2);
\draw (0,0)--++(2,0) arc (-90:90:2) --++(-2,0) arc (90:270:2);
\fill[white] (0,2)--++(2,0) arc (-90:90:.9) --++(-2,0) arc (90:270:.9);
\draw (0,2)--++(2,0) arc (-90:90:.9) --++(-2,0) arc (90:270:.9);
\foreach \y in {0,1}{
\foreach \x in {0,1,2}{ 
\fill[white] (\x,\y) to [bend right=30] (\x,{\y+1}) to [bend right=30] (\x,\y);
\draw (\x,\y) to [bend right=30] (\x,{\y+1}) to [bend right=30] (\x,\y);
}}
\foreach \y in {0,1,2}{
\foreach \x in {0,1,2}{
\fill[white] (\x,{\y-.2}) arc (270:-90:.2);
\draw[blue,thick,->] (\x,{\y-.2}) arc (270:-90:.2);
}}
\begin{scope}[shift={(1,2.9)}, scale=.5]
\node at (1.7,0) {0}; 
\fill[gray!50] (0,0) circle (1.1);
\fill[white] (0,0) circle (.9);
\draw (0,0) circle (1.1);
\draw (0,0) circle (.9);
\fill[white] (1,0) circle (.4);
\draw[red,thick] (1,0) circle (.4);
\draw[red,thick,->] (1.4,0) arc (0:360:.4);
\draw[blue] (-.5,.2) arc (-180:0:.5);
\draw[blue] (-.3,0) arc (180:0:.3);
\end{scope} 
\begin{scope}[shift={(0,-.4)}] 
\node at (0,0) {1};
\node at (0,1) {2};
\node at (0,2) {3};
\node at (1,0) {4};
\node at (1,1) {5};
\node at (1,2) {6};
\node at (2,0) {7};
\node at (2,1) {8};
\node at (2,2) {9};
\end{scope}
\end{tikzpicture}
\caption{Encoding Map for Shor's [[9,1,3]] Code}\label{Fig: 913}
\end{figure}

Shor also gave a QECC to correct arbitrary 1-qubit error in \cite{Shor95}. The encoding map is a composition of the encoding maps in \eqref{Equ: Encoding X}  in the $X$- and $Z$-basis. It encodes 1 logical qubit as 9 physical qubits, with distance 3, also called the [[9,1,3]] code. The corresponding diagrammatic representation is Fig.~\ref{Fig: 913}. Its generating stabilizers are given by cycles operators of the cycles around the eight contractible regions,
namely $X_1X_2$, $X_2X_3$, $X_4X_5$, $X_5X_6$, $X_7X_8$, $X_8X_9$ for the 6 shaded regions, and $Z_1Z_2Z_3Z_4Z_5Z_6$, $Z_4Z_5Z_6Z_7Z_8Z_9$ for the two unshaded regions. We consider such cycles to be {\it local}. In contrast, we consider the cycle around the hole to be {\it non-local}, and the corresponding cycle operator is $X_3X_6X_9$.
When the input, namely the logical qubit, is $\ket{0}_{X}$ or $\ket{1}_{X}$, we can remove the genus using the string-genus relation. Therefore two logical qubits can be represented by charged graphs without genus. 

\section{Toric Codes}
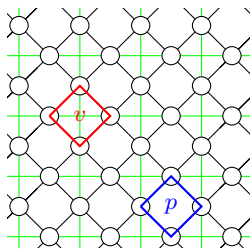
\begin{figure}[h]
\begin{tabular}{cc}
Periodic Square Lattice & Quantized Graph\\
\begin{tikzpicture}
\begin{scope}[scale=.8]
\foreach \x in {0,1,2,3}{
\draw (\x,-.2)--++(0,4);
}
\foreach \y in {0,1,2,3}{
\draw (-.2,\y)--++(4,0);
}
\node at (1,2) {$\textcolor{red}{v}$};
\node at (1-.5,2) {$\textcolor{red}{Z}$};
\node at (1+.5,2) {$\textcolor{red}{Z}$};
\node at (1,2-.5) {$\textcolor{red}{Z}$};
\node at (1,2+.5) {$\textcolor{red}{Z}$};
\node at (2.5,.5) {$\textcolor{blue}{p}$};
\node at (2.5-.5,.5) {$\textcolor{blue}{X}$};
\node at (2.5+.5,.5) {$\textcolor{blue}{X}$};
\node at (2.5,.5-.5) {$\textcolor{blue}{X}$};
\node at (2.5,.5+.5) {$\textcolor{blue}{X}$};
\end{scope}
\end{tikzpicture}
&
\begin{tikzpicture}
\begin{scope}[scale=.8]
\foreach \x in {0,1,2,3}{
\draw[green] (\x,-.2)--++(0,4);
}
\foreach \y in {0,1,2,3}{
\draw[green] (-.2,\y)--++(4,0);
}
\foreach \x in {0,1,2}{
\foreach \y in {0,1,2,3}{
\draw (\x+.3,\y-.2)--++(1,1);
\draw (\x-.2,\y+.7)--++(1,-1);
\draw (3+.3,\y-.2)--++(.5,.5);
\draw (0-.2,\y+.3)--++(.5,.5);
}}
\foreach \y in {1,2,3}{
\draw (3-.2,\y+.7)--++(1,-1);
}
\draw (3-.2,0+.7)--++(.9,-.9);
\foreach \x in {0,1,2,3}{
\draw (\x+.7,3.8)--++(.1,-.1);
}
\foreach \x in {0,1,2,3}{
\foreach \y in {0,1,2,3}{
\fill[white] (\x+.5,\y) circle (.15);
\draw (\x+.5,\y) circle (.15);
\fill[white] (\x,\y+.5) circle (.15);
\draw (\x,\y+.5) circle (.15);
}}
\fill[white] (0,-.5) rectangle (3.8,-.2);
\fill[white] (0,3.8) rectangle (3.8,4.3);
\fill[white] (-1,0) rectangle (-.2,3.8);
\fill[white] (3.8,0) rectangle (4.6,3.8);
\node at (1,2) {$\textcolor{red}{v}$};
\draw[red,thick](1-.5,2)--++(.5,-.5) --++(.5,.5)--++(-.5,.5)--++(-.5,-.5);
\node at (2.5,.5) {$\textcolor{blue}{p}$};
\draw[blue,thick](2.5-.5,.5)--++(.5,-.5) --++(.5,.5)--++(-.5,.5)--++(-.5,-.5);
\end{scope}
\end{tikzpicture}
\end{tabular}
\caption{Quantization of Square Lattices}\label{Fig: Lattice}
\end{figure}

\begin{figure}[h]
\begin{tikzpicture}
\begin{scope}[scale=.8]
\foreach \x in {0,1,2}{
\foreach \y in {0,1,2,3}{
\draw (\x+.3,\y-.2)--++(1,1);
\draw (\x-.2,\y+.7)--++(1,-1);
\draw (3+.3,\y-.2)--++(.5,.5);
\draw (0-.2,\y+.3)--++(.5,.5);
}}
\foreach \y in {1,2,3}{
\draw (3-.2,\y+.7)--++(1,-1);
}
\draw (3-.2,0+.7)--++(.9,-.9);
\foreach \x in {0,1,2,3}{
\draw (\x+.7,3.8)--++(.1,-.1);
}
\foreach \x in {0,1,2,3}{
\foreach \y in {0,1,2,3}{
\fill[white]  (\x+.5,\y-.15) arc (270:-90:.15);
\draw[blue,thick,->] (\x+.5,\y-.15) arc (270:-90:.15);
\fill[white] (\x-.15,\y+.5) arc (180:-180:.15);
\draw[blue,thick,->] (\x-.15,\y+.5) arc (180:-180:.15);
}}
\fill[white] (0,-.5) rectangle (3.8,-.2);
\fill[white] (0,3.8) rectangle (3.8,4.3);
\fill[white] (-1,0) rectangle (-.2,3.8);
\fill[white] (3.8,0) rectangle (4.6,3.8);
\foreach \x in {0}{
\foreach \y in {0,1,2,3}{
\draw (\x-.2,\y+.7) arc (90:270: {(.6*\y+1)/2}  and {.6*\y+1}) --++(5,0) arc (-90:90: {.6*\y+1}) --++(-1,0);
\draw (\x-.2,\y+.3) arc (90:270: {(.6*\y+.75)/2} and {.6*\y+.75}) --++(5,0) arc (-90:90: {.6*\y+.75}) --++(-1,0);
}}
\foreach \x in {0,1,2,3}{
\foreach \y in {0}{
\draw[white,WL] (\x+.7,\y-.2) --++(0,-1) arc (0:-180: {.6*\x+1+.5} and {(.6*\x+1+.5)/2}) --++(0,6) arc (180:0: {.6*\x+1+.5}) --++(0,-1);
\draw[white,WL] (\x+.3,\y-.2) --++(0,-1) arc (0:-180: {.6*\x+.75+.5} and {(.6*\x+.75+.5)/2} ) --++(0,6) arc (180:0: {.6*\x+.75+.5}) --++(0,-1);
\draw (\x+.7,\y-.2) --++(0,-1) arc (0:-180: {.6*\x+1+.5} and {(.6*\x+1+.5)/2}) --++(0,6) arc (180:0: {.6*\x+1+.5}) --++(0,-1);
\draw (\x+.3,\y-.2) --++(0,-1) arc (0:-180: {.6*\x+.75+.5} and {(.6*\x+.75+.5)/2} ) --++(0,6) arc (180:0: {.6*\x+.75+.5}) --++(0,-1);
}}
\begin{scope}[shift={(-1,4.5)}, scale=.5]
\node at (1.7,0) {0}; 
\fill[white] (0,0) circle (.9);
\draw (0,0) circle (1.1);
\draw (0,0) circle (.9);
\fill[white] (1,0) circle (.4);
\draw[red,thick] (1,0) circle (.4);
\draw[red,thick,->] (1.4,0) arc (0:360:.4);
\draw[blue] (-.5,.2) arc (-180:0:.5);
\draw[blue] (-.3,0) arc (180:0:.3);
\end{scope} 
\begin{scope}[shift={(4.5,-.5)}, scale=.5]
\node at (1.7,0) {1}; 
\fill[white] (0,0) circle (.9);
\draw (0,0) circle (1.1);
\draw (0,0) circle (.9);
\fill[white] (1,0) circle (.4);
\draw[red,thick] (1,0) circle (.4);
\draw[red,thick,->] (1.4,0) arc (0:360:.4);
\draw[blue] (-.5,.2) arc (-180:0:.5);
\draw[blue] (-.3,0) arc (180:0:.3);
\end{scope} 
\end{scope}
\end{tikzpicture}
\caption{Encoding Map of the Toric Code for $n=4$: The two discs marked by 0 and 1 are inputs. Their locations correspond to the two generators of the homological group $H_1\cong \mathbb{Z} \times \mathbb{Z}$ of the torus. The rest of the discs are outputs. We omit the alternating shadings of the lattice.}\label{Fig: Lattice Encode}
\end{figure}
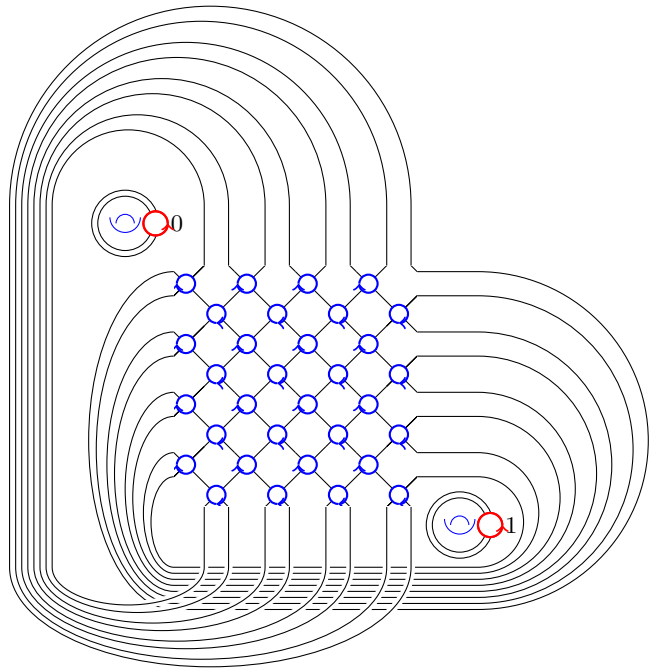

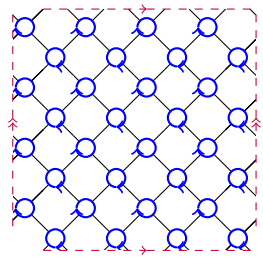
\begin{figure}[h]
\begin{tikzpicture}
\begin{scope}[scale=.8]
\foreach \x in {0,1,2}{
\foreach \y in {0,1,2,3}{
\draw (\x+.3,\y-.2)--++(1,1);
\draw (\x-.2,\y+.7)--++(1,-1);
\draw (3+.3,\y-.2)--++(.5,.5);
\draw (0-.2,\y+.3)--++(.5,.5);
}}
\foreach \y in {1,2,3}{
\draw (3-.2,\y+.7)--++(1,-1);
}
\draw (3-.2,0+.7)--++(.9,-.9);
\foreach \x in {0,1,2,3}{
\draw (\x+.7,3.8)--++(.1,-.1);
}
\foreach \x in {0,1,2,3}{
\foreach \y in {0,1,2,3}{
\fill[white]  (\x+.5,\y-.15) arc (270:-90:.15);
\draw[blue,thick,->] (\x+.5,\y-.15) arc (270:-90:.15);
\fill[white] (\x-.15,\y+.5) arc (180:-180:.15);
\draw[blue,thick,->] (\x-.15,\y+.5) arc (180:-180:.15);
}}
\fill[white] (0,-.5) rectangle (3.8,-.2);
\fill[white] (0,3.8) rectangle (3.8,4.3);
\fill[white] (-1,0) rectangle (-.2,3.8);
\fill[white] (3.8,0) rectangle (4.6,3.8);
\draw[purple,dashed,->>] (-.2,.2)--(-.2,2);
\draw[purple,dashed] (-.2,2)--(-.2,3.8);
\draw[purple,dashed,->>] (4-.2,.2)--(4-.2,2);
\draw[purple,dashed] (4-.2,2)--(4-.2,3.8);
\draw[purple,dashed,->] (.3,-.2)--(2,-.2);
\draw[purple,dashed] (2,-.2)--(3.8,-.2);
\draw[purple,dashed,->] (.3,4-.2)--(2,4-.2);
\draw[purple,dashed] (2,4-.2)--(3.8,4-.2);
\end{scope}
\end{tikzpicture}
\caption{Encoding Map of the Toric Code for $n=4$: 
Two pairs of boundary edges are glued, and each pair indicates an input. We plan to address the mathematical definition of the glued boundary in the companion paper \cite{Liu-Companion}.}\label{Fig: Lattice with boundary}
\end{figure}

Kitaev introduced his toric code to study topological order and topological quantum computation \cite{FKLW02,Kit03}.
The $[[n^2,2,n]]$ code is built on a $n\times n$ square lattice. Each edge has a qubit. The stabilizers are given by vertex operators $A(v)$ and plaquette operators $B(p)$, where $A(v)$ is the tensor product of Pauli $Z$'s on nearest qubits and $B(p)$ is the tensor product of Pauli $X$'s on nearest qubits. We illustrate the square lattice and its quantization in Fig.~\ref{Fig: Lattice}. All stabilizers becomes cycle operators of local cycles on the quantized graphs. The stabilizer states of these local operators encodes two logical qubits. We give this encoding map in Fig.~\ref{Fig: Lattice Encode} and an alternative illustration in Fig.~\ref{Fig: Lattice with boundary}.


In our pictorial approach, we can naturally capture topological orders, which are inconvenient to describe using graph states.
Kitaev represents the Pauli matrices as pairs of Majorana fermions to diagonalize the Hamiltonian given by the sum of vertex operators and plaquette operators. Our pictorial quon language naturally captures Kitaev's map and the eigenbasis of the Hamiltonian. 
Taking the two inputs to be $\ket{0}_{X}$ or $\ket{1}_{X}$, we can remove the genus using the string-genus relation.
We obtain four genus-0 charged graphs as an orthonormal basis of the ground state space.
They represent the topological orders of the toric code.
Moreover, we obtain all topological excitations by adding pairs of charges to the graphs.
Furthermore, we can construct new exactly solvable models in the Section {\it EXACTLY SOLVABLE MODELS}
extending this idea.

One can also consider a tessellation of a compact surface and construct a stabilizer code whose stabilizers are vertex operators and plaquette operators. Such codes have been considered as homological codes \cite{BomMar07}. We can obtain the encoding map of such homological codes by the quantization of the tessellation graph, similar to the toric code case.

In all cases above, the quantized graph has no braid. The stabilizers given by the local cycle operators are either the tensor product of $X$'s or the tensor product of $Z$'s. Stabilizer codes with such stabilizers are called CSS codes \cite{CS96,Ste96}.
The distance of the CSS code is the minimal of the distances of the two reduced classical error-correcting codes.

\begin{figure}[h]
\begin{tikzpicture}
\begin{scope}[scale=.8]
\foreach \x in {0,1,2}{
\foreach \y in {0,1,2,3}{
\draw (\x+.3,\y-.2)--++(1,1);
\draw (\x-.2,\y+.7)--++(1,-1);
\draw (3+.3,\y-.2)--++(.5,.5);
\draw (0-.2,\y+.3)--++(.5,.5);
}}
\foreach \y in {1,2,3}{
\draw (3-.2,\y+.7)--++(1,-1);
}
\draw (3-.2,0+.7)--++(.9,-.9);
\foreach \x in {0,1,2,3}{
\draw (\x+.7,3.8)--++(.1,-.1);
}
\foreach \x in {0,1,2,3}{
\foreach \y in {0,1,2,3}{
\fill[white]  (\x+.5,\y-.15) arc (270:-90:.15);
\draw[blue,thick,->] (\x+.5,\y-.15) arc (270:-90:.15);
\fill[white] (\x-.15,\y+.5) arc (180:-180:.15);
\draw[blue,thick,->] (\x-.15,\y+.5) arc (180:-180:.15);
}}
\fill[white] (0,-.5) rectangle (3.8,-.2);
\fill[white] (0,3.8) rectangle (3.8,4.3);
\fill[white] (-1,0) rectangle (-.2,3.8);
\fill[white] (3.8,0) rectangle (4.6,3.8);
\draw[purple,dashed,->>] (-.2,.2)--(-.2,2);
\draw[purple,dashed] (-.2,2)--(-.2,3.8);
\draw[purple,dashed,->>] (4-.2,.2)--(4-.2,2);
\draw[purple,dashed] (4-.2,2)--(4-.2,3.8);
\draw[purple,dashed,->] (.3,-.2)--(2,-.2);
\draw[purple,dashed] (2,-.2)--(3.8,-.2);
\draw[purple,dashed,->] (.3,4-.2)--(2,4-.2);
\draw[purple,dashed] (2,4-.2)--(3.8,4-.2);
\draw[white,line width=12pt,opacity=1.0] (.75,1.25)--(2.25,2.75);
\draw[purple] (.75,1.25)--(2.25,2.75);
\draw[orange,thick] (1.75,2.75)--++(-.25,.25)--++(.5,.5)--++(1,-1)--++(-.5,-.5)--++(-.25,.25);
\end{scope}
\end{tikzpicture}
\caption{Encoding Map of the Toric Code with Defects for $n=4$: 
The defect line is represented by the (purple) line on the top, and the original output discs are replaced by braids in the quantized graph.
The stabilizer at a defect endpoint is represented by a cycle operator of the (orange) cycle.}\label{Fig: Lattice with defects}
\end{figure}
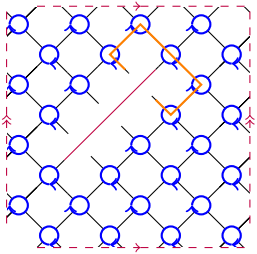

Kitaev and Kong studied the toric code with defect lines in \cite{KitKon12}. In our approach, we can represent a defect line as a line on a graph with braids as shown in Fig.~\ref{Fig: Lattice with defects}.
Moreover, the stabilizer at a defect endpoint (e.g. the stabilizer $Q$ in Equation (8) in \cite{KitKon12}) could also be represented as a cycle operator.

\section{graphical QECC}

\begin{figure}[h]
\begin{align*}
v_{\emptyset}&=
\raisebox{-1.5cm}{
\begin{tikzpicture}
\begin{scope}[scale=1]
\foreach \x in {1,2,3,4,5} 
{
\fill[white]  ({cos (\x*72)}, {sin (\x*72)}) circle (.1);
\draw  ({cos (\x*72)}, {sin (\x*72)}) circle (.1);
\node at ({1.2*cos (\x*72)}, {1.2*sin (\x*72)}) {\textcolor{black}{\x}};
}
\foreach \x in {0,1,2,3,4} 
{
\draw ({cos (\x*72)}, {sin (\x*72)})--({cos (\x*72+72)}, {sin (\x*72+72)});
\draw ({cos (\x*72)}, {sin (\x*72)})--({cos (\x*72+144)}, {sin (\x*72+144)});
}
\foreach \x in {0,1,2,3,4} 
{
\draw[white,WL] ({cos (\x*72)}, {sin (\x*72)})--({(cos (\x*72+144)+cos (\x*72))/2 }, {(sin (\x*72+144)+sin (\x*72))/2});
\draw ({cos (\x*72)}, {sin (\x*72)})--({(cos (\x*72+144)+cos (\x*72))/2 }, {(sin (\x*72+144)+sin (\x*72))/2});
}
\foreach \x in {1,2,3,4,5} 
{
\fill[white]  ({cos (\x*72)}, {sin (\x*72)}) circle (.1);
\draw[blue]  ({cos (\x*72)}, {sin (\x*72)}) circle (.1);
}
\end{scope}
\end{tikzpicture}}
&
v_{E(\Gamma)}&=
\raisebox{-1.5cm}{
\begin{tikzpicture}
\begin{scope}[scale=1]
\foreach \x in {0,1,2,3,4} 
{
\draw ({cos (\x*72)}, {sin (\x*72)})--({cos (\x*72+72)}, {sin (\x*72+72)});
\draw ({cos (\x*72)}, {sin (\x*72)})--({cos (\x*72+144)}, {sin (\x*72+144)});
}
\foreach \x in {0,1,2,3,4} 
{
\draw[white,WL] ({cos (\x*72)}, {sin (\x*72)})--({(cos (\x*72+144)+cos (\x*72))/2 }, {(sin (\x*72+144)+sin (\x*72))/2});
\draw ({cos (\x*72)}, {sin (\x*72)})--({(cos (\x*72+144)+cos (\x*72))/2 }, {(sin (\x*72+144)+sin (\x*72))/2});
}
\foreach \x in {1,2,3,4,5} 
{
\fill[white]  ({cos (\x*72)}, {sin (\x*72)}) circle (.1);
\draw[blue]  ({cos (\x*72)}, {sin (\x*72)}) circle (.1);
\node at ({1.2*cos (\x*72)}, {1.2*sin (\x*72)}) {\textcolor{black}{\x}};
}
\foreach \x in {1,2,3,4,5} 
{
\fill ({.5*(cos (\x*72)+cos (\x*72+72))}, {.5*(sin (\x*72)+sin (\x*72+72))}) circle (.05);
\fill ({.5*(cos (\x*72)+cos (\x*72+144))}, {.5*(sin (\x*72)+sin (\x*72+144))}) circle (.05);
}
\end{scope}
\end{tikzpicture}}
\end{align*}
\caption{Logical Qubits for the [[5,1,3]] Code: The five discs are output discs and the arrows are omitted. }\label{Fig: 513 logic}
\end{figure}
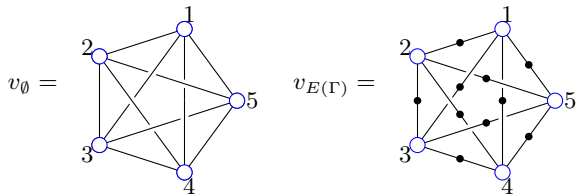

In this section, we give a new pictorial method to construct QECC. We compute the distance of such codes later.
First we construct a QECC, which is not a CSS code, from the smallest non planar graph $\Gamma=K_5$. We consider $K_5$ as a quantized graph without a pre-quantization.
Using the quon language, we construct two vectors in $(\mathbb{C}^2)^5$ from the two charged graphs in Fig.~\ref{Fig: 513 logic}.
They both have length $\sqrt{2}$ and they are orthogonal. So we identify them as logical qubits $\ket{0}_L$ and $\ket{1}_L$ after a global normalization by $\frac{1}{\sqrt{2}}$.
All their common stabilizers are given by cycles operators with even length. 
For example, the cycle operator of the cycle $L_1: 2\to 4\to 3 \to 5 \to 2$ is $O_{L_1}=X_2 Z_3 Z_4 X_5$.
The stabilizer group is generated by four stabilizers: 
\begin{align*}
X_2 Z_3 Z_4 X_5, \\
X_3 Z_4 Z_5 X_1, \\  
X_4 Z_5 Z_1 X_2, \\  
X_5 Z_1 Z_2 X_3, 
\end{align*}
corresponding to the four cycles
\begin{align*}
L_1: 2\to 4\to 3 \to 5 \to 2 \\
L_2: 3\to 5\to 4 \to 1 \to 3 \\
L_3: 4\to 1\to 5 \to 2 \to 4 \\
L_4: 5\to 2\to 1 \to 3 \to 5 \\
\end{align*}
The corresponding stabilizer code is known as the [[5,1,3]] code, first constructed in \cite{LMPZ96,BDSW96}.
It is the optimal QECC for one logical qubit and one qubit error.
We obtain the first pictorial interpretation of stabilizer group of the [[5,1,3]] code as even-length cycles on $K_5$.
We also obtain an internal permutation group $S_5$ symmetry of the [[5,1,3]] code, beyond the dihedral group $D_5$ symmetry of its graph state.


In general, suppose $\Gamma$ is a connected 4-valent graph with $n$ vertices. Let $E(\Gamma)$ be its edges, and $V(\Gamma)$ be its vertices.
Let $\mathcal{A}$ be the set of even subsets of $E(\Gamma)$ and $\mathcal{B}$ be the set of cycles of $\Gamma$. 
For any element $C \in \mathcal{A}$, we define a charged graph $\Gamma_C$, such that each edge in $C$ has a charge on $\Gamma$.
Using the quon language, the charged graph $\Gamma_C$ defines an $n$-qubit vector of length $\sqrt{2}$ in $(\mathbb{C}^2)^n$. We denote the corresponding vector state by $\phi_C$ after a global normalization by $\frac{1}{\sqrt{2}}$.

Take a subset $\mathcal{C}$ of $\mathcal{A}$, $\emptyset \in \mathcal{C}$,
we construct a {\it graphical QECC} $(\Gamma,\mathcal{C})$ as follows.
It is a stabilizer code whose stabilizer group $\mathcal{S}$ consists of cycle operators of cycles containing even charges on all charged graphs.
Technically, for a cycle $L$, we need to choose the sign $\pm$ for the cycle operator $O_L$, such that $O_L v_{\emptyset}=v_{\emptyset}.$ Such a choice ensures that $-I \notin \mathcal{S}$.

On the other hand, given a connected 4-valent graph $\Gamma$, and a set $\mathcal{L}$ of cycles, we can construct a {\it graphical QECC} $(\Gamma,\mathcal{L})$ whose stabilizer group is generated by the cycle operators $O_L$, $L \in \mathcal{L}$. Then the logical qubits are given by certain charged graphs, see the following sections for further explanations.

\section{Duality for Graphical QECC}\label{Sec: Duality}
We introduce a duality between the set $\mathcal{A}$ of all even subsets (for logical qubits) and the set $\mathcal{B}$ of all cycles (for stabilizers) of a connected 4-valent graph $\Gamma$.
Then both $\mathcal{A}$ and $\mathcal{B}$ can be regarded as subgroups of the additive group $2^{E(\Gamma)} \cong \ell^2(E(\Gamma),\mathbb{F}_2)$ of all subsets of $E(\Gamma)$. Moreover, the groups $\mathcal{A}$ and $\mathcal{B}$ are dual to each other with respect to the inner product on $\ell^2(E(\Gamma),\mathbb{F}_2)$.
We show that the two constructions of graphical QECC $(\Gamma,\mathcal{C})$ and $(\Gamma,\mathcal{L})$ are equivalent, see \cite{Liu-Companion} for a mathematical proof.

For an even subset $C\in \mathcal{A}$, we define
the characteristic function $1_C: E(\Gamma) \to \mathbb{F}_2$, taking value $1$ in $C$ and $0$ elsewhere.
We can consider the characteristic function $1_C$ as bits on the edges. Then we obtain an addition $C_1+C_2$ for the even subsets $C_1,C_2$ of $E(\Gamma)$, such that
$1_{C_1}+1_{C_2}=1_{C_1+C_2}$ over $\mathbb{F}_2$.

For a cycle $L \in \mathcal{B}$, we define the characteristic function $1_L : E(\Gamma) \to \mathbb{F}_2$, taking value $1$ on the edges in $L$ and $0$ elsewhere. We consider $1_L$ as bits on the edges.
For two cycles $L_1, L_2$, we define the cycle $L_1+L_2$, such that $1_{L_1}+1_{L_2}=1_{L_1+L_2}$ over $\mathbb{F}_2$.
Recall that $1_C$ is the characteristic function of an even subset $C$ of $E(\Gamma)$.
We say $C$ and $L$ are perpendicular, denoted by $C \perp L$, if $1_C \perp 1_L$ as functions in $\ell^2(E(\Gamma),\mathbb{F}_2)$.

Given $\mathcal{C} \subset \mathcal{A}$, the stabilizers of the graphical QECC $(\Gamma,\mathcal{C})$ are cycle operators 
$\mathcal{S}=\{O_L : L \in \mathcal{L}\}$, where $\mathcal{L}=\mathcal{C}^{\perp}:=\{ L \in \mathcal{B}:  L \perp C, ~\forall~ C\in \mathcal{C} \}$. The logical qubits are given by the superpositions of $\{\phi_C : C \in \overline{\mathcal{C}} \}$, where $\overline{\mathcal{C}}$ is the subgroup of $2^{E(\Gamma)}$ generated by $\mathcal{C}$. Moreover, 
\begin{align*}
\overline{\mathcal{C}}&=\mathcal{L}^{\perp}=\mathcal{C}^{\perp \perp}.
\end{align*}

On the other hand, given $\mathcal{L} \subset \mathcal{B}$, the stabilizers of the graphical QECC $(\Gamma,\mathcal{L})$ are cycle operators
$\mathcal{S}=\{O_L : L \in \overline{\mathcal{L}}\}$, where $\overline{\mathcal{L}}$ is the subgroup of $2^{E(\Gamma)}$ generated by $\mathcal{L}$. The logical qubits are given by the superpositions of $\{\phi_C : C \in \mathcal{C} \}$, where $\mathcal{C}=\mathcal{L}^{\perp}:=\{ C \in \mathcal{A}:  C \perp L, ~\forall~ L\in \mathcal{L} \}$. Moreover, 
\begin{align*}
\overline{\mathcal{L}}&=\mathcal{C}^{\perp}=\mathcal{L}^{\perp \perp}.
\end{align*}

As graphical QECC $(\Gamma,\mathcal{C})=(\Gamma,\overline{\mathcal{C}})$ and $(\Gamma,\mathcal{L})=(\Gamma,\overline{\mathcal{L}})$.
Moreover,  $(\Gamma,\mathcal{C})=(\Gamma,\mathcal{L})$, if $\overline{\mathcal{C}}=\mathcal{L}^{\perp}$ and  $\overline{\mathcal{L}}=\mathcal{C}^{\perp}$.
Therefore, the two constructions are equivalent up to this duality. 
We call the elements in $\overline{\mathcal{C}}= \mathcal{L}^{\perp}$ logical, as they corresponds to logical qubits.

%

\section{Bipartite Equivalence of Charged Graph States}
In this section, we introduce a \textit{bipartite} equivalence for charged graphs. The states of two charged graphs are same iff the difference of their charges is bipartite. Otherwise the states are orthogonal.

\begin{figure}[h]
\raisebox{-1.5cm}{
\begin{tikzpicture}
\begin{scope}[scale=1]
\foreach \x in {0,1,2,3,4} 
{
\draw ({cos (\x*72)}, {sin (\x*72)})--({cos (\x*72+72)}, {sin (\x*72+72)});
\draw ({cos (\x*72)}, {sin (\x*72)})--({cos (\x*72+144)}, {sin (\x*72+144)});
}
\foreach \x in {1,2,3,4,5} 
{
\fill[white]  ({cos (\x*72)}, {sin (\x*72)}) circle (.1);
\draw  ({cos (\x*72)}, {sin (\x*72)}) circle (.1);
}
\foreach \x in {1,3} 
{
\fill ({.5*(cos (\x*72)+cos (\x*72+72))}, {.5*(sin (\x*72)+sin (\x*72+72))}) circle (.05);
}
\foreach \x in {1,2,3,5} 
{
\fill ({.5*(cos (\x*72)+cos (\x*72+144))}, {.5*(sin (\x*72)+sin (\x*72+144))}) circle (.05);
}
\foreach \x in {2,4,5} 
{
}
\foreach \x in {4} 
{
}
\end{scope}
\end{tikzpicture}}
\caption{A bipartite set $C_0$ of charges on $K_5$: We omit the choices of the braids in the graph, because they will not affect the definitions and the results in this section.}\label{Fig: bipartite}
\end{figure}
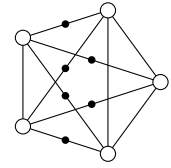

Given a 4-valent connected graph $\Gamma$, we call an even subset $C\in \mathcal{A}$ bipartite, if there is a partition of the vertices $V(\Gamma)=V_1 \sqcup V_2$, 
such that $C$ is the set of edges in $E(\Gamma)$ connecting $V_1$ and $V_2$. 
We prove that $\phi_{C_1}=\phi_{C_2}$ iff $C_1+C_2$ is bipartite, otherwise $\phi_{C_1}\perp \phi_{C_2}$ in \cite{Liu-Companion}.
For example, the set of charges $C_0$ in Fig.~\ref{Fig: bipartite} is bipartite.
In particular, $\phi_{C_0}=\phi_{\emptyset}=\ket{0}_L$ for the [[5,1,3]] code.

Let $B(\Gamma)$ be the set of bipartite subsets of $E(\Gamma)$. Then $B(\Gamma)$ forms an additive subgroup of $\mathcal{B}$. For an even subset $C$, we denote $[C]=C+B(\Gamma)$ to be the equivalence class of $C$ modulo the bipartite relation. 
Then the states $\{\phi_{C} : C \in \mathcal{A}\}$ modulo bipartite relations form an orthonormal basis of the space of physical qubits. We call it the Fourier basis of $\Gamma$, denoted by $B_{\mathfrak{F}}(\Gamma)$.

\section{Quantum Linear Systems}
Recall that logical bits in a classical QEC can be described as the solutions of a linear system, as shown in Equation \eqref{Equ: linear system} for the repetition code. 
In this section, we describe the logical qubits of a graphical QECC as the solutions of a quantum linear system using the quantized graph. 

Given a finite 4-valent connected graph $\Gamma$, and a subset $\mathcal{L}$ of $\mathcal{B}$ cycles on the graph, we obtain a graphical QECC $(\Gamma,\mathcal{L})$. An orthonormal basis of the logical qubits are given by the states $\{\phi_{C} : C\in \mathcal{C}=\mathcal{L}^{\perp} \}$, modulo the bipartite equivalence.
Therefore, we can solve $C\in \mathcal{C}$ as the solution of the linear system over $\mathbb{F}_2$ consisting of linear equations $C \perp L$, for any $L \in \mathcal{L}$.
Moreover, the logical qubits are superpositions of such $\Phi_{C}$, $C\in \mathcal{C}=\mathcal{L}^{\perp}$. Therefore, we call $C \perp L$ a quantum linear system of the graphical QECC, and 
we consider the logical qubits of the graphical QECC as the solution of this quantum linear system.

%
%

For example, the underling graph of the [[5,1,3]] code is $\Gamma=K_5$ as shown in Fig.~\ref{Fig: 513 logic}. 
We denote the edge connecting vertices $i$ and $j$ in the graph $\Gamma$ by $e_{ij}$.
Let $E(\Gamma)$ be the set of edges of the graph $\Gamma=K_5$.
For an even subset $C$ of $E(\Gamma)$, we take $c_{ij}=1_{C}(e_{ij})$ in $\mathbb{F}_2$.
Recall that the generating stabilizers of the [[5,1,3]] code are defined by the cycle operators $O_{L_k}$ of the four cycles $\mathcal{L}=\{L_k : k=1,2,3,4\}$. We describe the logical qubits using the following linear system over $\mathbb{F}_2$: 
The state $\phi_C$ is a logic qubit, namely it is invariant under the action of the four cycle operators, iff
the following linear system over $\mathbb{F}_2$ holds: 
\begin{align*}
c_{24}+c_{43}+c_{35}+c_{52}=0 ;\\
c_{35}+c_{54}+c_{41}+c_{13}=0 ;\\
c_{41}+c_{15}+c_{52}+c_{24}=0 ;\\
c_{52}+c_{21}+c_{13}+c_{35}=0 .
\end{align*}
In other words, each of the four cycles contains even charges in $\Gamma_{C}$.
Moreover, any logical qubit is a superposition of the states of such charged graphs.
We consider the above linear system as a quantum linear system of the [[5,1,3]] code. Its solutions (modulo the bipartite equivalence) form an orthonormal basis of the logical qubits. 

We can write the coefficient matrix as
\begin{align}\label{Equ: quantum linear system}
A=\left[
\begin{array}{cccccccccc}
0 & 0 & 1 & 0 & 0 & 0 & 1 & 1 & 0 & 1  \\
0 & 0 & 0 & 1 & 0 & 1 & 0 & 1 & 1 & 0  \\
0 & 0 & 0 & 0 & 1 & 0 & 1 & 0 & 1& 1  \\
1 & 0 & 0 & 0 & 0 & 1 & 0 & 1 & 0 & 1  
\end{array}
\right],
\end{align}
when the coordinates are ordered as $e_{12},e_{23},e_{34},e_{45},e_{51},e_{13},e_{24},e_{35},e_{41},e_{52}$.
This is a quantum analogue of the matrix in Equation \eqref{Equ: linear system} for ECC.

\section{Distance}
In this section, we compute the distance of the graphical QECC in terms of its graphical data, see \cite{Liu-Companion} for the mathematical proof.
Suppose $\Gamma$ is a 4-valent connected graph. 
Due to the duality between $\mathcal{A}$ and $\mathcal{B}$ in the Section {\it DUALITY FOR GRAPHICAL QECC}, 
we take
$\mathcal{C}\subset \mathcal{A}$ and $\mathcal{L}\subset \mathcal{B}$, such that 
\begin{align*}
\mathcal{C}&=\overline{\mathcal{C}}=\mathcal{L}^{\perp}; \\
\mathcal{L}&=\overline{\mathcal{L}}=\mathcal{C}^{\perp}.
\end{align*}
Then $(\Gamma, \mathcal{C})$ are $(\Gamma, \mathcal{L})$ are the same graphical QECC.

For any even subset $C \in \mathcal{A}$, we define its weight $w(C)$ to be the minimal $m$, such that $C$ is the sum of $m$ pairs of adjacent edges.
We define the bipartite weight of $C$ as 
\begin{align*}
w_b(C) &= \min_{C' \in C+B(\Gamma)} w(C') .
\end{align*}
Then $w_b(C)=0$ iff $C$ is bipartite.
For even subsets $C_1$ and $C_2$ of $E(\Gamma)$, we define their relative bipartite weight as
\begin{align*}
w_b(C_1,C_2)&=w_b(C_1+C_2).\\
\end{align*}
Then $w_b$ is a distance function on the bipartite equivalence classes of even subsets of $E(\Gamma)$.
We define the bipartite weight of $\mathcal{C}$ as
\begin{align*}
w_b(\mathcal{C})=\min \{w_b(C) : C \in \mathcal{C}, w_b(C)\neq 0 \} \;.
\end{align*}
For any cycle $L \in \mathcal{B}$, we define its essential length $\ell(L)$ as the number of degree-two vertices of $L$.
We define the essential length of $\mathcal{L}$ as
\begin{align*}
\ell(\mathcal{L})=\min \{\ell(L) : L \in \mathcal{B}\setminus\mathcal{L} \} \;.
\end{align*}
The distance $d$ of the graphical QECC $(\Gamma, \mathcal{C})=(\Gamma, \mathcal{L})$ is given by
\begin{align}\label{Equ: distance}
d=\min\{w_b(\mathcal{C}), \ell(\mathcal{L})\}.
\end{align}
It is worth mentioning that computing the distance $d$ of a graphical QECC reduces to two independent graphical problems, namely computing $w_b(\mathcal{C})$ and $\ell(\mathcal{L})$. The only constraint between $\mathcal{C}$ and $\mathcal{L}$ is $\mathcal{C} \perp \mathcal{L}$.

\section{Fundamental Theorems}
For any connected 4-valent graph $\Gamma$, we call an [[n,k,d]] graphical QECC on $\Gamma$ optimal, if there is no graphical QECC on $\Gamma$ with a larger $k$ or $d$.
We characterize all optimal [[n,k,d]] graphical QECC on $\Gamma$ using Equation \eqref{Equ: distance}.
We plan to address the proofs in the companion paper \cite{Liu-Companion}.

For any $D\in \mathbb{N}$, we define
\begin{align*}
\mathcal{C}_{D}&:=\{C: w_b(C)\geq D\} \; ; \\
\mathcal{L}_{D}&:=\{L \in \mathcal{B} : \ell(L)<D \} \; ; \\
\mathcal{I}_{D}&:=\mathcal{C}_{D} \cap \mathcal{L_{D}}^{\perp} \; .\\
\end{align*}
Note that when $D$ increases, $\mathcal{C}_{D}$, $\mathcal{L}_{D}$, $\mathcal{I}_{D}$ decrease.
We define the code distance of $\Gamma$ as 
\begin{align}\label{Equ: Code Distance}
d(\Gamma):= \max\{ D:  \mathcal{I}_{D} \neq \emptyset \}.
\end{align}
We have the following optimal theorem: The maximal distance to construct an $[[n,1,d]]$ graphical QECC on $\Gamma$ is $d=d(\Gamma)$.

Note that if $C\in \mathcal{I}_{D}$, then $[C]\subseteq \mathcal{I}_{D}$.
We define $k_{\Gamma}(D)$ to be the maximal number, such that $\mathcal{C}/B(\Gamma) \cong \mathbb{Z}_2^{k_{\Gamma}(D)}$ for some group $\mathcal{C} \subseteq \mathcal{I}_{D} \cup B(\Gamma)$. We call such a group $\mathcal{C}$ optimal.
Then $k_{\Gamma}(D)$ is a decreasing function, and we call it the optimal function of $\Gamma$.

We have the following theorem for optimal $k$: For any $D$, $k=k_{\Gamma}(D)$ is the maximum to construct an $[[n,k,d]]$ graphical QECC on $\Gamma$, such that $d\geq D$.

As $k_{\Gamma}(D)$ is decreasing, we can take $1\leq d_1<d_2<\ldots<d_m = d(\Gamma)$, such that 
\begin{align*}
k_{\Gamma}(d_1)&=k_{\Gamma}(1); \\
k_{\Gamma}(d_j) &> k_{\Gamma}(d_j+1)=k_{\Gamma}(d_{j+1}), ~\forall~ 1 \leq j \leq m-1. \\
\end{align*}
Then we have the following theorem for optimal graphical QECC on $\Gamma$:
For any $1 \leq j \leq m$ and any optimal $\mathcal{C} \subseteq \mathcal{I}_{d_j} \cup B(\Gamma)$, $(\Gamma,\mathcal{C})$ is an $[[n,k_{\Gamma}(d_j), d_j ]]$ optimal graphical QECC $(\Gamma,\mathcal{C})$. Conversely, any optimal graphical QECC on $\Gamma$ arises in this way.

By a quick estimate, we have the following existence theorem:
For any 4-valent connected graph $\Gamma$, we can construct an $[[n,k,d]]$ graphical QECC, such that
\begin{itemize}
\item[(1)] $n$ is the number of vertices in $\Gamma$;
\item[(2)] $d \geq \ell(\Gamma)$;
\item[(3)] $\displaystyle k\geq n - \log_2 \sum_{j=0}^{d-1} C_{n}^j 3^j$,
\end{itemize}
where $\ell(\Gamma)=\ell(\emptyset)$ is the minimal essential length of cycles on $\Gamma$, and we call it the {\it essential girth} of $\Gamma$. 
The inequality (3) is similar to the quantum Gilbert-Varshamov bound \cite{ABKL00}, but we do not assume the non-degenerate condition.

In general, it could be difficult to construct a graph with small essential girth. We can improve our results by eliminating the cycles with small essential length.
For any $D>0$,  we have that $\overline{\mathcal{L}_{D}} \cong \mathbb{Z}_2^{s}$.
We can construct an $[[n,k,d]]$ graphical QECC, such that
\begin{itemize}
\item[(1)] $n$ is the number of vertices in $\Gamma$;
\item[(2)] $d \geq D$;
\item[(3)] $\displaystyle k\geq n-s - \log_2 \sum_{j=0}^{d-1} C_{n}^j 3^j$,
\end{itemize}
whenever $n-s - \log_2 \sum_{j=0}^{d-1} C_{n}^j 3^j \geq 1$.

We summarize the basic properties of the graphical QECC in Table~\ref{Table: graphical QECC}, extending Table.~\ref{Table: Quantization} for ECC.

\begin{table}[h]
\begin{tabular}{c|c}
connected 4-valent vertex graph $\Gamma$ & Graphical QECC\\
\hline
4-valent vertex (disc) & physical qubit \\
cycle $L$ & stabilizer $O_L$ \\
charged set $C$ & logical bit $\phi_C$
\end{tabular}
\caption{graphical QECC}\label{Table: graphical QECC}
\end{table}

\section{M\"{o}bius Codes}

In this section, we construct a family of graphical QECC on the M\"{o}bius Strip, which we call M\"{o}bius codes.
For each $n \in \mathbb{N}$, we construct an [[n(2n-1), 1, n]] QECC. 
It is similar to the toric code locally, but not globally. The distance can be computed as the minimal length of non-contractible loops on the pairs of pre-quantized graphs, see the mathematical proof in the companion paper \cite{Liu-Companion}.
We plan to compute their threshold in the future.

\begin{figure}[h]
\begin{tikzpicture}
\begin{scope}[scale=.8]
\foreach \z in {1,2,3,4}{
\draw[white,WL] ({-\z},\z) rectangle ({-\z+5},{\z-5});
\draw ({-\z},\z) rectangle ({-\z+5},{\z-5});
}
\foreach \x in {1}{
\foreach \y in {1,3}{
\fill[white]  (\x,\y-.15) arc (270:-90:.15);
\draw[blue,thick,->] (\x,\y-.15) arc (270:-90:.15);
}}
\foreach \x in {1}{
\foreach \y in {2}{
\fill[white]  (\x,\y-.15) arc (270:-90:.15);
\draw[blue,thick,->] (\x-.15,\y) arc (180:-180:.15);
}}
\foreach \x in {2}{
\foreach \y in {2}{
\fill[white]  (\x,\y-.15) arc (270:-90:.15);
\draw[blue,thick,->] (\x,\y-.15) arc (270:-90:.15);
}}
\foreach \x in {2}{
\foreach \y in {1}{
\fill[white]  (\x,\y-.15) arc (270:-90:.15);
\draw[blue,thick,->] (\x-.15,\y) arc (180:-180:.15);
}}
\foreach \x in {3}{
\foreach \y in {1}{
\fill[white]  (\x,\y-.15) arc (270:-90:.15);
\draw[blue,thick,->] (\x,\y-.15) arc (270:-90:.15);
}}
\node[blue] at (1+.3,1-.3) {1};
\node[blue] at (1-.3,2-.3) {2};
\node[blue] at (1+.3,3-.3) {3};
\node[blue] at (2-.3,1-.3) {4};
\node[blue] at (2+.3,2-.3) {5};
\node[blue] at (3+.3,1-.3) {6};
\begin{scope}[shift={(0,0)}, scale=.5]
\node at (1.7,0) {0}; 
\fill[white] (0,0) circle (.9);
\draw (0,0) circle (1.1);
\draw (0,0) circle (.9);
\fill[white] (1,0) circle (.4);
\draw[red,thick] (1,0) circle (.4);
\draw[red,thick,->] (1.4,0) arc (0:360:.4);
\draw[blue] (-.5,.2) arc (-180:0:.5);
\draw[blue] (-.3,0) arc (180:0:.3);
\end{scope} 
\end{scope}
\end{tikzpicture}
\caption{Encoding Map of the [[n(2n-1),1, n]] M\"{o}bius Code for $n=2$: It has one input disc labelled by 0, and six output discs labelled by 1-6.}\label{Fig: Mobius 2}
\end{figure}
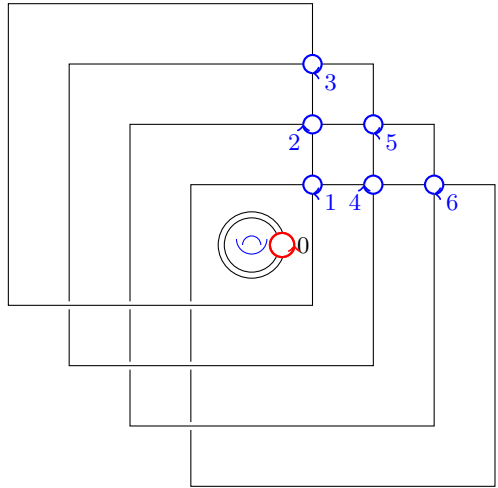

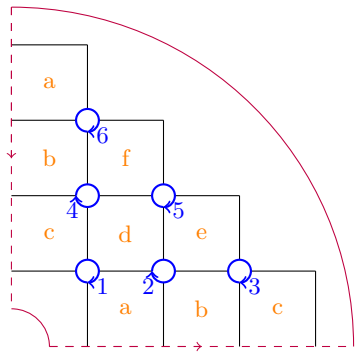
\begin{figure}[h]
\begin{tikzpicture}
\node[blue] at (1+.2,1-.2) {1};
\node[blue] at (2-.2,1-.2) {2};
\node[blue] at (3+.2,1-.2) {3};
\node[blue] at (1-.2,2-.2) {4};
\node[blue] at (2+.2,2-.2) {5};
\node[blue] at (1+.2,3-.2) {6};
\node[orange] at (1+.5,1-.5) {a};
\node[orange] at (2+.5,1-.5) {b};
\node[orange] at (3+.5,1-.5) {c};
\node[orange] at (1+.5,2-.5) {d};
\node[orange] at (2+.5,2-.5) {e};
\node[orange] at (1+.5,3-.5) {f};
\node[orange] at (1-.5,3+.5) {a};
\node[orange] at (1-.5,2+.5) {b};
\node[orange] at (1-.5,1+.5) {c};
\foreach \x in {1,2,3,4}{
\draw (\x,0)--(\x,{5-\x});
}
\foreach \y in {1,2,3,4}{
\draw (0,\y)--({5-\y},\y);
}
\foreach \x in {1}{
\foreach \y in {1,3}{
\fill[white]  (\x,\y-.15) arc (270:-90:.15);
\draw[blue,thick,->] (\x,\y-.15) arc (270:-90:.15);
}}
\foreach \x in {1}{
\foreach \y in {2}{
\fill[white]  (\x,\y-.15) arc (270:-90:.15);
\draw[blue,thick,->] (\x-.15,\y) arc (180:-180:.15);
}}
\foreach \x in {2}{
\foreach \y in {2}{
\fill[white]  (\x,\y-.15) arc (270:-90:.15);
\draw[blue,thick,->] (\x,\y-.15) arc (270:-90:.15);
}}
\foreach \x in {2}{
\foreach \y in {1}{
\fill[white]  (\x,\y-.15) arc (270:-90:.15);
\draw[blue,thick,->] (\x-.15,\y) arc (180:-180:.15);
}}
\foreach \x in {3}{
\foreach \y in {1}{
\fill[white]  (\x,\y-.15) arc (270:-90:.15);
\draw[blue,thick,->] (\x,\y-.15) arc (270:-90:.15);
}}
\draw[purple,dashed,->] (.5,0)--(2.5,0);
\draw[purple,dashed] (.5,0)--(4.5,0);
\draw[purple,dashed,->] (0,4.5)--(0,2.5);
\draw[purple,dashed] (0,4.5)--(0,.5);
\draw[purple] (.5,0) arc (0:90:.5);
\draw[purple] (4.5,0) arc (0:90:4.5);
\end{tikzpicture}
\caption{An alternative illustration on the M\"{o}bius strip: The M\"{o}bius strip is obtained by gluing a pair of edges (dashed lines with arrows) on the boundary. The graphical QECC in Fig.~\ref{Fig: Mobius 2} is essentially defined on M\"{o}bius strip.}\label{Fig: Mobius strip}
\end{figure}

The encoding maps for $n=2$ is illustrated in Fig.~\ref{Fig: Mobius 2}.  It is more conceptual to visualize this QECC on a M\"{o}bius strip, see Fig.~\ref{Fig: Mobius strip}. It has six contractible regions $a ,b,\ldots,f$ on the M\"{o}bius strip, which define six cycle operators as stabilizers: 
\begin{align*}
O_a &= X_1 X_2 X_6;\\
O_b &= Z_2 Z_3 Z_4 Z_6;\\
O_c &= X_1 X_3 X_4;\\
O_d &= Z_1 Z_2 Z_4 Z_6;\\
O_e &= X_2 X_3 X_5;\\
O_f &= X_4 X_5 X_6 .\\
\end{align*}
They generate a stabilizer group $(\mathbb{Z}_2)^5$ with one conservation law $O_aO_cO_eO_f=1$.
This construction can be generalized to arbitrary size, see Fig.~\ref{Fig: Mobius 3} for the case $n=3$.

\begin{figure}[h]
\begin{tikzpicture}
\begin{scope}[scale=.4]
\foreach \z in {1,2,3,4}{
\draw[white,WL] ({-\z},\z) rectangle ({-\z+5},{\z-5});
\draw ({-\z},\z) rectangle ({-\z+5},{\z-5});
}
\foreach \x in {1}{
\foreach \y in {1,3}{
\fill[white]  (\x,\y-.15) arc (270:-90:.15);
\draw[blue,thick,] (\x,\y-.15) arc (270:-90:.15);
}}
\foreach \x in {1}{
\foreach \y in {2}{
\fill[white]  (\x,\y-.15) arc (270:-90:.15);
\draw[blue,thick,] (\x-.15,\y) arc (180:-180:.15);
}}
\foreach \x in {2}{
\foreach \y in {2}{
\fill[white]  (\x,\y-.15) arc (270:-90:.15);
\draw[blue,thick,] (\x,\y-.15) arc (270:-90:.15);
}}
\foreach \x in {2}{
\foreach \y in {1}{
\fill[white]  (\x,\y-.15) arc (270:-90:.15);
\draw[blue,thick,] (\x-.15,\y) arc (180:-180:.15);
}}
\foreach \x in {3}{
\foreach \y in {1}{
\fill[white]  (\x,\y-.15) arc (270:-90:.15);
\draw[blue,thick,] (\x,\y-.15) arc (270:-90:.15);
}}
\end{scope}
\begin{scope}[scale=.4, shift={(10,0)}]
\foreach \z in {1,2,3,4}{
\draw[white,WL] ({-\z},\z) rectangle ({-\z+5},{\z-5});
\draw ({-\z},\z) rectangle ({-\z+5},{\z-5});
}
\foreach \x in {1}{
\foreach \y in {1,3}{
\fill[white]  (\x,\y-.15) arc (270:-90:.15);
\draw[blue,thick,] (\x,\y-.15) arc (270:-90:.15);
}}
\foreach \x in {1}{
\foreach \y in {2}{
\fill[white]  (\x,\y-.15) arc (270:-90:.15);
\draw[blue,thick,] (\x-.15,\y) arc (180:-180:.15);
}}
\foreach \x in {2}{
\foreach \y in {2}{
\fill[white]  (\x,\y-.15) arc (270:-90:.15);
\draw[blue,thick,] (\x,\y-.15) arc (270:-90:.15);
}}
\foreach \x in {2}{
\foreach \y in {1}{
\fill[white]  (\x,\y-.15) arc (270:-90:.15);
\draw[blue,thick,] (\x-.15,\y) arc (180:-180:.15);
}}
\foreach \x in {3}{
\foreach \y in {1}{
\fill[white]  (\x,\y-.15) arc (270:-90:.15);
\draw[blue,thick,] (\x,\y-.15) arc (270:-90:.15);
}}
\foreach \x in {1,2,3,4}{
\fill (\x,0) circle (.15);
}
\end{scope}
\end{tikzpicture}
\caption{Charged Graphs for Logical Qubits: We omit the labels and orientations of the output discs in both graphs.}\label{Fig: Logic 2}
\end{figure}
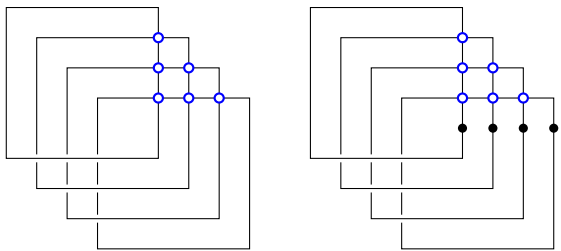

The quantized graph $\Gamma$ of this graphical QECC is given by the connected component with output discs. The logical qubits $\ket{0}_L$ and $\ket{1}_L$ can be described using charged graphs as illustrated in Fig.~\ref{Fig: Logic 2}. 

\begin{figure}[h]
\begin{tikzpicture}
\begin{scope}[scale=.4]
\foreach \z in {1,2,3,4,5,6}{
\draw[white,WL] ({-\z},\z) rectangle ({-\z+7},{\z-7});
\draw ({-\z},\z) rectangle ({-\z+7},{\z-7});
}
\foreach \x in {1}{
\foreach \y in {1,2,3,4,5}{
\fill[white]  (\x,\y-.15) arc (270:-90:.15);
\draw[blue,thick] (\x,\y-.15) arc (270:-90:.15);
}}
\foreach \x in {2}{
\foreach \y in {1,2,3,4}{
\fill[white]  (\x,\y-.15) arc (270:-90:.15);
\draw[blue,thick] (\x,\y-.15) arc (270:-90:.15);
}}
\foreach \x in {3}{
\foreach \y in {1,2,3}{
\fill[white]  (\x,\y-.15) arc (270:-90:.15);
\draw[blue,thick] (\x,\y-.15) arc (270:-90:.15);
}}
\foreach \x in {4}{
\foreach \y in {1,2}{
\fill[white]  (\x,\y-.15) arc (270:-90:.15);
\draw[blue,thick] (\x,\y-.15) arc (270:-90:.15);
}}
\foreach \x in {5}{
\foreach \y in {1}{
\fill[white]  (\x,\y-.15) arc (270:-90:.15);
\draw[blue,thick] (\x,\y-.15) arc (270:-90:.15);
}}
\begin{scope}[shift={(0,0)}, scale=.5]
\fill[white] (0,0) circle (.9);
\draw (0,0) circle (1.1);
\draw (0,0) circle (.9);
\fill[white] (1,0) circle (.4);
\draw[red,thick] (1,0) circle (.4);
\draw[red,thick,] (1.4,0) arc (0:360:.4);
\draw[blue] (-.5,.2) arc (-180:0:.5);
\draw[blue] (-.3,0) arc (180:0:.3);
\end{scope} 
\end{scope}
\end{tikzpicture}
\caption{Encoding Map of the [[n(2n-1),1, n]] M\"{o}bius Code for $n=3$: We omit the labels and orientations of the input/output discs.}\label{Fig: Mobius 3}
\end{figure}
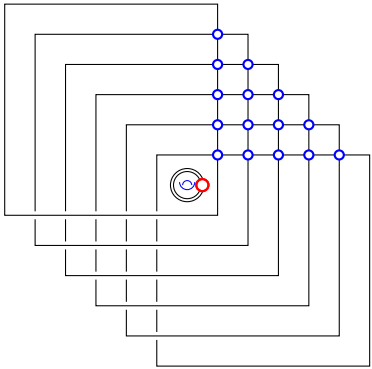

\section{Klein Codes}

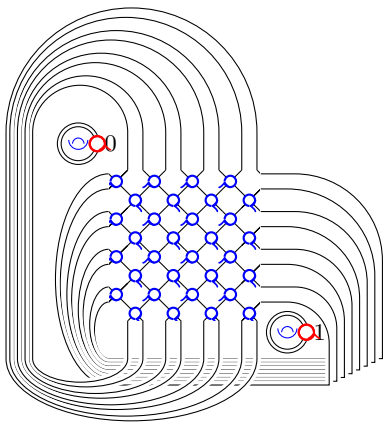
\begin{figure}[h]
\begin{tikzpicture}
\begin{scope}[scale=.5]
\foreach \x in {0,1,2}{
\foreach \y in {0,1,2,3}{
\draw (\x+.3,\y-.2)--++(1,1);
\draw (\x-.2,\y+.7)--++(1,-1);
\draw (3+.3,\y-.2)--++(.5,.5);
\draw (0-.2,\y+.3)--++(.5,.5);
}}
\foreach \y in {1,2,3}{
\draw (3-.2,\y+.7)--++(1,-1);
}
\draw (3-.2,0+.7)--++(.9,-.9);
\foreach \x in {0,1,2,3}{
\draw (\x+.7,3.8)--++(.1,-.1);
}
\foreach \x in {0,1,2,3}{
\foreach \y in {0,1,2,3}{
\fill[white]  (\x+.5,\y-.15) arc (270:-90:.15);
\draw[blue,thick,->] (\x+.5,\y-.15) arc (270:-90:.15);
\fill[white] (\x-.15,\y+.5) arc (180:-180:.15);
\draw[blue,thick,->] (\x-.15,\y+.5) arc (180:-180:.15);
}}
\fill[white] (0,-.5) rectangle (3.8,-.2);
\fill[white] (0,3.8) rectangle (3.8,4.3);
\fill[white] (-1,0) rectangle (-.2,3.8);
\fill[white] (3.8,0) rectangle (4.6,3.8);
\foreach \x in {0}{
\foreach \y in {0,1,2,3}{
\draw[white,WL] (\x-.2,\y+.3) arc (90:270: {(.6*\y+.75)/2} and {.6*\y+.75}) --++(5,0) --++({2.2-.4*\y},0)--++(0,{2.6-.4*\y}) arc (0:90: {-.4*\y+2.3}) --++(-.9,0);
\draw (\x-.2,\y+.3) arc (90:270: {(.6*\y+.75)/2} and {.6*\y+.75}) --++(5,0) --++({2.2-.4*\y},0)--++(0,{2.6-.4*\y}) arc (0:90: {-.4*\y+2.3}) --++(-.9,0);
\draw[white,WL] (\x-.2,\y+.7) arc (90:270: {(.6*\y+1)/2}  and {.6*\y+1}) --++(5,0) --++({2-.4*\y},0)--++(0,{2.3-.4*\y}) arc (0:90: {-.4*\y+2.3}) --++(-.7,0);
\draw (\x-.2,\y+.7) arc (90:270: {(.6*\y+1)/2}  and {.6*\y+1}) --++(5,0) --++({2-.4*\y},0)--++(0,{2.3-.4*\y}) arc (0:90: {-.4*\y+2.3}) --++(-.7,0);
}}
\foreach \x in {0,1,2,3}{
\foreach \y in {0}{
\draw[white,WL] (\x+.7,\y-.2) --++(0,-1) arc (0:-180: {.6*\x+1+.5} and {(.6*\x+1+.5)/2}) --++(0,6) arc (180:0: {.6*\x+1+.5}) --++(0,-1);
\draw[white,WL] (\x+.3,\y-.2) --++(0,-1) arc (0:-180: {.6*\x+.75+.5} and {(.6*\x+.75+.5)/2} ) --++(0,6) arc (180:0: {.6*\x+.75+.5}) --++(0,-1);
\draw (\x+.7,\y-.2) --++(0,-1) arc (0:-180: {.6*\x+1+.5} and {(.6*\x+1+.5)/2}) --++(0,6) arc (180:0: {.6*\x+1+.5}) --++(0,-1);
\draw (\x+.3,\y-.2) --++(0,-1) arc (0:-180: {.6*\x+.75+.5} and {(.6*\x+.75+.5)/2} ) --++(0,6) arc (180:0: {.6*\x+.75+.5}) --++(0,-1);
}}
\begin{scope}[shift={(-1,4.5)}, scale=.5]
\node at (1.7,0) {0}; 
\fill[white] (0,0) circle (.9);
\draw (0,0) circle (1.1);
\draw (0,0) circle (.9);
\fill[white] (1,0) circle (.4);
\draw[red,thick] (1,0) circle (.4);
\draw[red,thick,->] (1.4,0) arc (0:360:.4);
\draw[blue] (-.5,.2) arc (-180:0:.5);
\draw[blue] (-.3,0) arc (180:0:.3);
\end{scope} 
\begin{scope}[shift={(4.5,-.5)}, scale=.5]
\node at (1.7,0) {1}; 
\fill[white] (0,0) circle (.9);
\draw (0,0) circle (1.1);
\draw (0,0) circle (.9);
\fill[white] (1,0) circle (.4);
\draw[red,thick] (1,0) circle (.4);
\draw[red,thick,->] (1.4,0) arc (0:360:.4);
\draw[blue] (-.5,.2) arc (-180:0:.5);
\draw[blue] (-.3,0) arc (180:0:.3);
\end{scope} 
\end{scope}
\end{tikzpicture}
\caption{Encoding Map of the Klein Code: The two discs marked by 0 and 1 are input. Their locations correspond to the two generators of the homological group $H_1\cong \mathbb{Z} \times \mathbb{Z}$ of the torus. The rest discs are output. We omit the alternating shadings of the lattice.}\label{Fig: Klein Code}
\end{figure}

\begin{figure}[h]
\begin{tikzpicture}
\begin{scope}[scale=.8]
\foreach \x in {0,1,2}{
\foreach \y in {0,1,2,3}{
\draw (\x+.3,\y-.2)--++(1,1);
\draw (\x-.2,\y+.7)--++(1,-1);
\draw (3+.3,\y-.2)--++(.5,.5);
\draw (0-.2,\y+.3)--++(.5,.5);
}}
\foreach \y in {1,2,3}{
\draw (3-.2,\y+.7)--++(1,-1);
}
\draw (3-.2,0+.7)--++(.9,-.9);
\foreach \x in {0,1,2,3}{
\draw (\x+.7,3.8)--++(.1,-.1);
}
\foreach \x in {0,1,2,3}{
\foreach \y in {0,1,2,3}{
\fill[white]  (\x+.5,\y-.15) arc (270:-90:.15);
\draw[blue,thick,->] (\x+.5,\y-.15) arc (270:-90:.15);
\fill[white] (\x-.15,\y+.5) arc (180:-180:.15);
\draw[blue,thick,->] (\x-.15,\y+.5) arc (180:-180:.15);
}}
\fill[white] (0,-.5) rectangle (3.8,-.2);
\fill[white] (0,3.8) rectangle (3.8,4.3);
\fill[white] (-1,0) rectangle (-.2,3.8);
\fill[white] (3.8,0) rectangle (4.6,3.8);
\draw[purple,dashed,->>] (-.2,3.8)--(-.2,2);
\draw[purple,dashed] (-.2,.2)--(-.2,3.8);
\draw[purple,dashed,->>] (4-.2,.2)--(4-.2,2);
\draw[purple,dashed] (4-.2,2)--(4-.2,3.8);
\draw[purple,dashed,->] (.3,-.2)--(2,-.2);
\draw[purple,dashed] (2,-.2)--(3.8,-.2);
\draw[purple,dashed,->] (.3,4-.2)--(2,4-.2);
\draw[purple,dashed] (2,4-.2)--(3.8,4-.2);
\end{scope}
\end{tikzpicture}
\caption{Encoding Map of the Klein Code for $n=4$: 
Two pairs of edges are glued, and each pair indicates an input.}\label{Fig: Klein Lattice}
\end{figure}
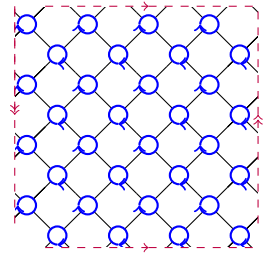

In this section, we construct a family of graphical QECC on the Klein bottle, which we call Klein codes.
For each $n \in \mathbb{N}$, we construct an $[[2n^2, 2, n]]$ QECC. The encoding map is given in Fig.\ref{Fig: Klein Code}, and an alternative notation is given in Fig.~\ref{Fig: Klein Lattice}. 
Its stabilizers are given by contractible cycle operators, similar to the M\"{o}bius codes. The distance can be computed as the minimal length of non-contractible loops on the pairs of pre-quantized graphs, see the mathematical proof in the companion paper \cite{Liu-Companion}.
We will compute their threshold in the future.

\section{Group Theory and QECC}

Given a group $\tilde{G}$ presented by two generators $a,b$ modulo relations, 
$$\tilde{G}=<a^{\pm1},b^{\pm1} : \sim >,$$ 
its Cayley graph is a 4-valent connected graph $\tilde{\Gamma}$ described in Fig.~\ref{Fig: Cayley Graph}.
For example, the Cayley graph of the free group $F_2$ with two generators is a 4-valent tree.

\begin{figure}[h]
\begin{tabular}{c|c}
$\tilde{G}$   & $\tilde{\Gamma}$ \\
\hline
element & vertex  \\
$g^{-1}h \in \{a^{\pm1},b^{\pm1}\}$ & edge $(g,h)$ \\
relation & cycle \\
word length & path length 
\end{tabular}
\caption{Cayley Graph}\label{Fig: Cayley Graph}
\end{figure}

Let $G$ be a finite quotient of $\tilde{G}$. We construct a graphical QECC using the data $(\tilde{G},G,a,b)$.
Let $\pi: \tilde{G} \to G$ be the quotient map.
Let $\tilde{\Gamma}$, $\Gamma$ be the Cayley graphs of $\tilde{G}$ and $G$ w.r.t. the generators $a,b$.
Let $\tilde{\Gamma}$ be the Cayley graph of $\tilde{G}$ w.r.t. the generators $a,b$.
The induced quotient map on the Cayley graphs is still denoted by $\pi : \tilde{\Gamma} \to \Gamma$.
For each cycle $L$ in $\tilde{\Gamma}$, we call $\pi(L)$ a {\it contractible cycle} in $\Gamma$.
Let $\mathcal{L}$ be the set of all contractible cycles in $\Gamma$. Then $\overline{\mathcal{L}}=\mathcal{L}$.
We obtained a graphical QECC $(\Gamma,\mathcal{L})$, denoted by $(\tilde{G},G,a,b)_g$ in terms of group data.

Example: Shor's quantum repetition code in Fig.~\ref{Fig: QRC Quon} could be recovered by $(\tilde{G},G,a,b)_g$, where
\begin{enumerate}
\item $\tilde{G}=\mathbb{Z}_6$;
\item $G=\mathbb{Z}_3$;
\item $a=b=1$.
\end{enumerate}

Example: Kitaev's toric code is given by
\begin{enumerate}
\item $\tilde{G}=\{(x,y) : x,y \in \mathbb{Z}, x+y \in 2\mathbb{Z} \}$,
\item $G=\{(x,y) : x,y \in \mathbb{Z}_n, x+y \in 2\mathbb{Z} \}$,
\item $a=(1,1), b=(-1,1)$.
\end{enumerate}

Example: Wen's toric code \cite{Wen03} is given by 
\begin{enumerate}
\item $\tilde{G}=\mathbb{Z}^2$,
\item $G=\mathbb{Z}_n^2$,
\item $a=(1,0), b=(0,1)$.
\end{enumerate}

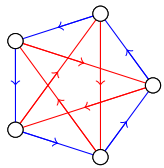
\begin{figure}[h]
\raisebox{-1.5cm}{
\begin{tikzpicture}
\begin{scope}[scale=1]
\foreach \x in {1,2,3,4,5} 
{
\fill[white]  ({cos (\x*72)}, {sin (\x*72)}) circle (.1);
\draw  ({cos (\x*72)}, {sin (\x*72)}) circle (.1);
}
\foreach \x in {0,1,2,3,4} 
{
\draw[blue] ({cos (\x*72)}, {sin (\x*72)})--({cos (\x*72+72)}, {sin (\x*72+72)});
\draw[blue,->] ({cos (\x*72)}, {sin (\x*72)})--({.5*(cos (\x*72+72)+cos (\x*72))}, {.5*(sin (\x*72+72)+sin (\x*72))});
\draw[red] ({cos (\x*72)}, {sin (\x*72)})--({cos (\x*72+144)}, {sin (\x*72+144)});
\draw[red,->] ({cos (\x*72+144)}, {sin (\x*72+144)})--({.5*(cos (\x*72+144)+cos (\x*72))}, {.5*(sin (\x*72+144)+sin (\x*72))});
}
\foreach \x in {1,2,3,4,5} 
{
\fill[white]  ({cos (\x*72)}, {sin (\x*72)}) circle (.1);
\draw  ({cos (\x*72)}, {sin (\x*72)}) circle (.1);
}
\end{scope}
\end{tikzpicture}}
\caption{Cayley Graph $K_5$: The edges of the pentagon and of the star correspond to the generators $a$ and $b$ respectively.}\label{Fig: Cayley Graph K5}
\end{figure}

Example: The [[5,1,3]] code is given by
\begin{enumerate}
\item $\tilde{G}=\Z_{10}$,
\item  $G=\Z_{3}$, 
\item  $\textcolor{blue}{a=1}$, $\textcolor{red}{b=3}$,
\end{enumerate}
The graph $\Gamma=K_5$ has been used in the graphical construction of the $[[5,1,3]]$ code. Now it arises from the Cayley graph of $G$ as illustrated in Fig.~\ref{Fig: Cayley Graph K5}.
The Cayley Graph $\tilde{\Gamma}$ is a double cover of $\Gamma$. The cycles in $\tilde{\Gamma}$ always have even length, so the contractible cycles in $\Gamma$ have even length. We obtain the stabilizers of the $[[5,1,3]]$ code from these contractible cycles characterized by $\tilde{G}$.

For the toric code, the Cayley graph $\Gamma$ of $G$ is the quantized graph in Fig.~\ref{Fig: Lattice}. The Cayley graph $\tilde{\Gamma}$ of $\tilde{G}$ is naturally embedded in the plane, which can be considered as the universal cover of $\Gamma$. All cycles on $\Gamma$ are contractible. We use their image in $\Gamma$ to describe the contractible cycles in $\Gamma$, without introducing the underlying torus. The logical qubits are encoded by non-contractible cycles on $\Gamma$, namely topological orders.

In general, the Cayley graph $\Gamma$ may not be embedded in a compact surface. Topological features of such graphical QECC on $\Gamma$ may not be apparent. 
Our interpretation of the contractible cycles on $\Gamma$ using $\tilde{G}$ is a natural generalization for the graphical QECC $(\tilde{G},G,a,b)_g$.

\subsection{Geometric Groups and QECC}
Several known QECC with geometric properties could be unified by the graphical QECC from the geometric group $\tilde{G}=<a,b : a^{p}, b^{q}, (ab)^{r}>$.
The Cayley graph $\tilde{\Gamma}$ can be canonically embedded in a surface $M$ with constant Gaussian curvature $\kappa$, such that
\begin{align*}
\kappa&=1, && 1/p+1/q+1/r >1; \\
\kappa&=0, && 1/p+1/q+1/r =1; \\
\kappa&=-1, && 1/p+1/q+1/r <1.  
\end{align*}
Moreover, $M$ is a sphere, a plane, or a Poincare discs for the three cases.
Precisely, the surface $M$ has a tessellation of isomorphic triangles $\Delta ABC$, such that the triangle has angles $\frac{\pi}{p}$, $\frac{\pi}{q}$, $\frac{\pi}{r}$ at $A$, $B$, $C$ respectively. It is called a Poincare triangle, when $\kappa<0$.
Let $\Psi$ be the graph on $M$, consisting of the vertices $A$, $B$ and the edge $AB$ of all triangles.
Then the Cayley graph $\tilde{\Gamma}$ is the quantized graph of $\Psi$.
We obtain a graphical QECC from any finite quotient $G$ of $\tilde{G}$.

When $\kappa=1$, $\tilde{G}$ is a finite group, and the parameters has an $ADE$ classification. We obtain finitely many QECCs.

When $\kappa=0$, $\tilde{G}$ is an infinite group, the solutions of $1/p+1/q+1/r =1$ are $(2,2,\infty)$, $(2,3,6)$, $(2,4,4)$, and $(3,3,3)$ up to a permutation.
The Cayley graph $\Gamma$ of a finite quotient $G$ is embedded in a torus. We obtain a graphical QECC $(\tilde{G},G, a, b)_g$ on a torus with different tessellations for different $G$. For example, when $(p,q,r)=(3,3,3)$, we obtain QECC on the Honeycomb lattice.

When $\kappa=-1$, $\tilde{G}$ is an infinite group, and there are infinitely many solutions of $1/p+1/q+1/r <1$.
The Cayley graph $\Gamma$ of a finite quotient $G$ is embedded in a higher genus surface with curvature $-1$.
The QECC for $p=2,q=3,r=7$ has been studied as QECC on Hurwitz surfaces in \cite{Kim07}.
The QECC for $p=q$ has been studied as a hyperbolic surface QECC in \cite{BreTer16}.

\section{LDPC QECC}
Given a connected 4-valent graph $\gamma$, and a set of of cycles $\mathcal{L} \subset \mathcal{B}$, we constructed a graphical QECC $(\gamma,\mathcal{L})$. We also obtained a quantum linear system whose solutions are logical qubits of the code. Its coefficient matrix $A$ is an $m \times 2n$ matrix with over $\mathbb{F}_2$, where $m$ is the number of cycles in $\mathcal{L}$. See an example in Equation \ref{Equ: quantum linear system} for the [[5,1,3]] code. We call $(\gamma,\mathcal{L})$ a low-density parity-check (LDPC) code, if $A$ is sparse, namely the proportion of 1's in every row and every column of $A$ is small. This is equivalent to the notion of LDPC stabilizer code, which we plan to prove in the companion paper \cite{Liu-Companion}. In this section, we assume that $\tilde{G}$ is represented by the generators $a,b$ with $k$ relations. Let $R$ be the maximal length of the relations. Usually $R \ll n$.
We show that the graphical QECC $(G,\tilde{G},a,b)$ is a LDPC code.

For any group element $g \in G$ and a relation $r$ of $\tilde{G}$, we obtain a cycle $\tilde{L}(g,r)$ starting at $g$ moving along $r$ in the Cayley graph $\tilde{\Gamma}$.
As the size of the cycle in $\tilde{\Gamma}$ is bounded by the length of the relation $r$, we call such cycle {\it local}.
Let $L(g,r)$ be the image of $\tilde{L}(g,r)$ in $\Gamma$.
We call the cycle $L(g,r)$ and the corresponding cycle operator $O_{L(g,r)}$ local as well. 
Let $\mathcal{L}=\{L(g,r): \forall g, r\}$ be the set of $kn$ local cycles on $\Gamma$.
Any contractible cycle operator on $\Gamma$ is generated by these local cycle operator, because any cycle on the Cayley graph $\tilde{\Gamma}$ is a sum of local cycles (mod 2).
Therefore $\overline{\mathcal{L}}$ is the set of contractible cycles.
Therefore $(G,\tilde{G},a,b)=(\Gamma,\overline{\mathcal{L}})=(\Gamma,\mathcal{L})$ as a QECC.
Let $A$ be the $kn \times 2n $ coefficient matrix of the quantum linear system of $(\Gamma,\mathcal{L})$. The number of 1's in every row is bounded by $R$. 
The number of $1's$ in every column is bounded by $kR$. The proportion is bounded by $\frac{R}{n}$. As $R \ll n$, $(\Gamma,\mathcal{L})$ is LDPC.
One can also check it directly that the Pauli $X,Z$ stabilizer matrix of $(\Gamma,\mathcal{L})$ is sparse, and $(\Gamma,\mathcal{L})$ is a LDPC stabilizer code.

\section{Exactly Solvable Models}\label{Sec: Exactly Solvable Models}
It is important to implement the space of logical qubits as the ground states of a local Hamiltonian $H$, which is more accessible in the laboratory. 
We give a natural construction of such a translation-invariant local Hamiltonian for the graphical QECC $(\tilde{G}, G, a, b)$, when $\tilde{G}$ is presented by finitely many relations. 

Suppose $\tilde{G}$ is represented by the generators $a,b$ with finitely many relations. For any $g \in G$ and any relation $r$, we obtain a local operator $O_{L(g,r)}$ defined in the Section \textit{LDPC QECC}. Its diameter on the Cayley graph $\Gamma$ is bounded by $1/2$ the length of $r$. 

For example, in the toric code, the group $\tilde{G}=\{(x,y) : x,y \in \mathbb{Z}, x+y \in 2\mathbb{Z} \}$ can be represented as $\tilde{G}=<a,b : aba^{-1}b^{-1} >$.
Moving along the relation $aba^{-1}b^{-1}$ stating at $g \in \tilde{G}$, we obtain a length-4 cycle. The corresponding cycle operator is either a vertex operator $A(v)$ or a plaquette operator $B(p)$. Take all $g \in G$, we obtain all vertex operators and plaquette operators, which are considered to be local operators in the toric code. The Hamiltonian $H$ is the opposite of the sum of these local operators.  Then the logical qubits are given by the ground states of $H$. We define the local Hamiltonian to be 
\begin{align*}
H=-\sum_{g, r} O_{L(g,r)}. 
\end{align*} 
When we change $g$, the cycle moves transitively on the Cayley graph $\Gamma$. 
So the Hamiltonian $H$ is translation invariant w.r.t. the group action on the Cayley graph.
The ground states of $H$ are the logical qubits of $(\tilde{G}, G, a, b)$, because the local operators generate the stabilizer group. 
Using the duality in the Section \textit{DUALITY FOR GRAPHICAL QECC}, an orthonormal basis the ground states is given by states of charged graphs $\phi_{C}$, $C \in \mathcal{L}^{\perp}$ modulo the bipartite equivalence.
Moreover, an eigenbasis of the Hamiltonian is given by states of charged graphs $\phi_{C}$ for $C \in \mathcal{A}$ modulo the bipartite equivalence. Furthermore, the partition function can be expressed in terms of the graphical data of $\Gamma$. Therefore, we obtained an exactly solvable model from $(\tilde{G}, G, a, b)$ and a finite presentation of $\tilde{G}$.

Suppose $\tilde{G}$ is an infinite group generated by $a,b$ with finitely many relations. We can also define an exactly solvable model on the infinite Cayley graph $\Gamma$.  
The Hilbert space of ground states and excitations on the infinite Cayley can also be rigorously defined using the Gelfand-Naimark-Segal construction.
By taking a net of finite quotients $\{G_n\}$ of $\tilde{G}$, we can consider the model on $\Gamma$ as a large scale limit of the models $(\tilde{G}, G_n, a, b)$ on the quotients.
The Hamiltonian is gapped in the large scale limit.

For example, in Wen's toric code, we can take
$\tilde{G}=\mathbb{Z}^2=<a,b : aba^{-1}b^{-1}>$, 
$a=(1,0), b=(0,1)$,
$G_n=\{(x,y) : x,y \in \mathbb{Z}_n, \}$.
This provides an approximation of the infinite square lattice by periodic finite lattices, such that their Hamiltonians share the same local operators defined by the relation $aba^{-1}b^{-1}$. The relation $aba^{-1}b^{-1}$ corresponds to the square cycle operators on the lattice.

\section{Concluding Remarks}
In this paper, we establish a quantization process for graphs and apply the quantized graphs to construct graphical QECC. Several properties of the graphical QECC are described by the graphical data of the quantized graph. 
The code has further group symmetries if the underlying graph is the Cayley graph of a group. In this case, we obtain a low-density parity-check QECC. 
Moreover, its logical qubits can be implemented by the ground states of a translation-invariant local Hamiltonian of an exactly solvable model. If the group is infinite, then the Hamiltonian is gapped in a large scale limit.
It would be interesting to implement these ground states as a topological quantum field theory (TQFT). 
 
The connected condition for the quantized graph is a technical assumption to simplify the statements. From the view of the quon language, one can generalize the quantized graphs in two different directions: Firstly, one can study quantized graphs in a higher-genus surface. Then one would obtain all stabilizer codes; On the other hand, one can replace the braids in the quantized graph by other phase transformations. Then one could obtain the encoding map for non-stabilizer codes. In principal, any encoding map could be obtained by combining the two generalizations. We expect to discover new internal structures by investigating these two generalizations of our methods.

\section{Acknowledgement}

The author would like to thank Arthur Jaffe for his constant encouragement and hospitality, and to thank Xun Gao, Sirui Lu, Yunxiang Ren, Alina Vdovina, Xiaogang Wen and Youwei Zhao for helpful discussions. The author thanks Harvard University for hospitality. The author was supported by Grant 100301004 from Tsinghua University and by Templeton Religion Trust under the grant TRT 159. The author presented an early version of the results at the conference Current Progress in Mathematical Physics in December 2018 at Harvard University.

\end{document}